\documentclass[
reprint,
amsmath,
amssymb,
aps,
floatfix,
]{revtex4-2}

\usepackage{ifxetex}
\ifxetex
\usepackage{xeCJK}
\else
\usepackage{CJKutf8}
\fi

\usepackage{graphicx}
\usepackage{dcolumn}
\usepackage{bm}
\usepackage{nccmath}
\usepackage{amsmath}
\usepackage{empheq}
\usepackage{subfigure}
\usepackage{makecell}
\usepackage{comment}
\usepackage{multirow}
\usepackage{tabularx}
\usepackage{booktabs}
\usepackage{float}
\usepackage{hyperref}
\usepackage{array}

\newcommand{\mi}{\mathrm{i}}

\usepackage{siunitx}

\begin{document}
	
	%\preprint{APS/123-QED}
	
	\title{Black holes immersed in modified Chaplygin-like dark fluid and cloud of strings: shadows, quasinomal modes and greybody factors}%
	
	\author{Hao-Peng Yan$^{a}$} \email{Corresponding author: yanhaopeng@tyut.edu.cn}
	\author{Zeng-Yi Zhang$^{a}$}
	\author{Xiao-Jun Yue$^{a}$} 
	\author{Xiang-Qian Li$^{a}$} \email{Corresponding author: lixiangqian@tyut.edu.cn}
	\affiliation{$^{a}$ College of Physics and Optoelectronic Engineering, Taiyuan University of Technology, Taiyuan 030024, China}

\begin{abstract}
We present a unified investigation of black hole shadows, quasinormal modes (QNMs), and greybody factors (GBFs) for a static, spherically symmetric black hole within a composite environment of a modified Chaplygin-like dark fluid (MCDF) and a cloud of strings (CoS). We examine the structure of critical photon orbits and the corresponding optical appearance under spherical accretion. Using the Wentzel–Kramers–Brillouin (WKB) approximation, we compute the quasinormal frequencies and greybody spectra, and explore their correspondence with the black hole shadows in the eikonal limit. A systematic parameter study demonstrates that the CoS intensity has the primary influence on the shadows, QNMs and GBFs, while the MCDF parameters introduce more complex but characterizable modifications to each. Our results demonstrate that these environmental components imprint distinct yet interrelated signatures on key observables, offering specific predictions for probing exotic black hole environments.
\end{abstract}

\maketitle

\section{Introduction}
\label{sec:intro}
The existence of black holes, once a purely theoretical prediction of General Relativity, is now firmly established by astronomical observations. The landmark images of the supermassive black holes M87* and Sgr A* obtained by the Event Horizon Telescope collaboration have inaugurated a new era of strong-field gravity astronomy, wherein the black hole shadow provides a direct probe of spacetime geometry in the immediate vicinity of the event horizon \cite{EventHorizonTelescope:2019dse, EventHorizonTelescope:2022wkp}. Concurrently, the advent of gravitational wave astronomy has opened a complementary observational window into the dynamics of compact objects, with the ringdown phase characterized by quasinormal modes (QNMs)—complex-frequency oscillations that constitute unique gravitational wave fingerprints of black holes \cite{Abbott:2016blz,Cardoso:2017cqb}.

These observational breakthroughs demand a deeper theoretical understanding of black holes in environments that more realistically reflect their cosmological context. To this end, three principal theoretical tools have emerged as crucial diagnostics of black hole structure and dynamics. Black hole shadows characterize the optical appearance through critical photon orbits \cite{Perlick:2021aok, Kocherlakota:2022jnz, Hou:2022eev, Hou:2024qqo, Guo:2018kis, Li:2020drn, Hu:2020usx, Zhao:2025ouq, Fan:2024rsa, Gui:2024owz, Zeng:2020dco, Zeng:2021dlj, Zeng:2023zlf, He:2025hbu, He:2025qmq, Wang:2025ihg,Battista:2023iyu,Wang:2025fmz,Tsukamoto:2014tja,Tsukamoto:2017fxq}. Recent explorations have expanded this framework to diverse environments \cite{Chen:2005qh,Konoplya:2019sns,Zeng:2025nmu,Zeng:2022fdm,Zeng:2025kqw}, with contemporary research even investigating the intriguing possibility of using shadows as novel probes for particle dark matter \cite{Chen:2024nua}. QNMs reveal the stability and characteristic ringing of spacetime under external perturbations \cite{Berti:2009kk,Konoplya:2011qq, Teukolsky:1973ha,Press:1973zz,Nollert:1999ji,Chen:2025sbz,Guo:2024jhg,Liu:2024bfj, Zhidenko:2003wq}. Greybody factors (GBFs) govern the transmission probability of Hawking radiation to distant observers, providing essential insights into semi-classical evaporation processes and quantum effects in curved spacetime \cite{Hawking:1975vc,Decanini:2011xi, Zhang:2020qam}. 

Furthermore, recent studies have revealed profound physical connections among these three diagnostics, establishing a rigorous framework for probing strong-field gravity. 
In the eikonal limit, a well-established correspondence links QNM frequencies to the properties of the photon sphere and consequently to the shadow radius~\cite{Cardoso:2008bp,Yang:2012he,Jusufi:2019ltj, Li:2021zct,Campos:2021sff, Konoplya:2025mvj, Xiong:2025wgs}. 
Complementing this geometric picture, recent research has demonstrated that the greybody factor plays a pivotal role in governing the excitation amplitude of the ringdown signal, thereby directly linking transmission probabilities to gravitational wave observables~\cite{Oshita:2021iyn}. 
Concurrently, observations suggest that greybody spectra often exhibit greater robustness than quasinormal frequencies under small perturbations of the geometry, offering a complementary diagnostic for the near-horizon structure~\cite{Rosato:2024arw,Oshita:2024fzf}. 
More broadly, the scattering interpretation has been deepened to explicitly connect quasinormal frequencies with the transmission and reflection coefficients of the effective potential~\cite{Konoplya:2024lir,Konoplya:2019hlu}. 
This multifaceted correspondence between geometry (shadow), dynamics (QNMs), and scattering (GBFs) now serves as a concrete paradigm for synergistically analyzing black hole physics across different observational channels~\cite{ Pedrotti:2025idg, Guo:2022hjp,Guo:2021wid,Guo:2020nci,Shi:2025gst}.

The application of these tools becomes particularly significant in the context of modern cosmology, where dark energy and dark matter dominate the energy content \cite{SupernovaSearchTeam:1998fmf,SupernovaCosmologyProject:1998vns}. Among various proposals, quintessence dark energy offers a compelling explanation for cosmic acceleration \cite{Wang:1999fa,Bahcall:1999xn}. The foundational work by Kiselev, who first derived a static, spherically symmetric black hole solution surrounded by quintessence matter \cite{Kiselev:2002dx}, established a framework for studying how dark energy influences black hole shadows \cite{Lacroix:2012nz, Khan:2020ngg,  Zeng:2020vsj, He:2021aeo, Zeng:2025kqw}, QNMs and GBFs \cite{Chen:2005qh,Gogoi:2023lvw,Fatima:2024gji,Hamil:2024njs}. Unified models that simultaneously describe dark matter and dark energy have attracted significant attention, among which the Chaplygin gas \cite{Kamenshchik:2001cp} and its generalizations stand as prominent candidates. These models successfully account for diverse cosmological observations \cite{Kamenshchik:2001cp, Bilic:2001cg, Bento:2002ps, Park:2009np,Aggarwal:2025sqz}. Notably, the Chaplygin gas emerges naturally within string theory frameworks, transcending its purely phenomenological origins \cite{Ogawa:2000gj,Bordemann:1993ep,Jackiw:2000cc}. Black holes immersed in Chaplygin-like dark fluids—including the original Chaplygin-like dark fluid (CDF), its generalized extension (GCDF), and the modified version (MCDF)—have been extensively studied \cite{Li:2019lhr, Li:2019ndh, Li:2022csn, Li:2023zfl,Li:2024abk,Li:2025ftb,Sekhmani:2024rpf,Becar:2024agj,Bohra:2025oro}.

Further enriching this landscape, the cloud of strings (CoS) model provides a minimalist geometric framework for describing how one-dimensional extended objects can modify black hole spacetimes. 
Originally developed by Letelier~\cite{Letelier:1979ej}, this model generalizes the concept of a dust cloud to a collection of one-dimensional strings. 
Physically, it can be interpreted as a macroscopic effective description of topological defects---analogous to cosmic strings---formed during symmetry-breaking phase transitions in the early universe. 
While not intended to model standard astrophysical accretion flows (such as gas or dust), the CoS serves as a valuable toy model for investigating the gravitational imprints of global anisotropy and extended structures. 
By treating the background as a continuous fluid with string-like stress, this framework allows one to isolate and study how remnant networks of primordial defects might modify the classical observables in strong-field gravity, distinct from point-particle matter distributions.
Recently, studies of black holes immersed in a CoS have attracted growing interest~\cite{He:2021aeo,Zeng:2022fdm,Hamil:2024nrv,Sadeghi:2019muh,Sadeghi:2020bsa,Sadeghi:2023tzf}.

Building upon previous work \cite{Li:2025ftb}, which investigated timelike and null geodesics, shadows, and images of MCDF-CoS black holes surrounded by thin accretion disks, the present study extends this analysis to encompass the QNM spectrum and GBFs, while systematically exploring the fundamental relationships among the black hole shadow, QNMs, and GBFs within this framework.

This paper is structured as follows: Section~\ref{sec:blackholebackground} introduces the static, spherically symmetric black hole solution immersed in both MCDF and CoS. Section~\ref{sec:shadow_appearance} analyzes critical photon orbits and the black hole shadow, alongside optical appearances under spherical accretion. Section~\ref{sec:qnm} presents QNM computations and discusses their physical implications. Section~\ref{sec:gbf} examines GBFs and their parameter dependence. Finally, Section~\ref{conclusion} synthesizes our principal findings and discusses their broader implications.

\begin{figure*}[]
\centering % \begin{center}/\end{center} takes some additional vertical space
\includegraphics[width=.394\textwidth]{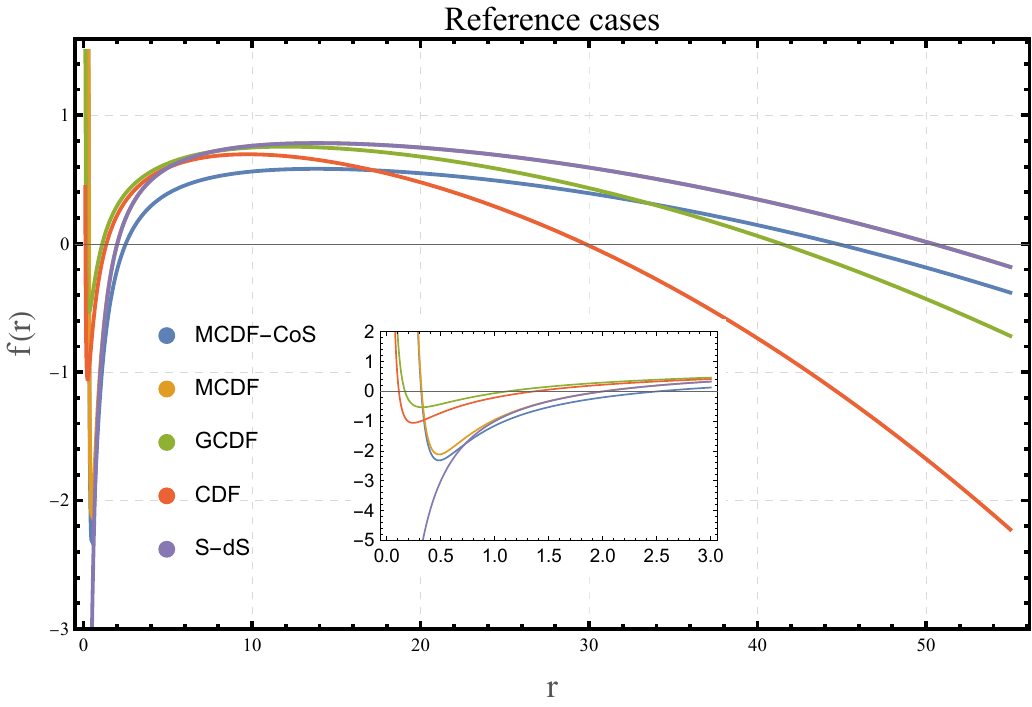}\hspace{0.05\textwidth}
\includegraphics[width=.394\textwidth]{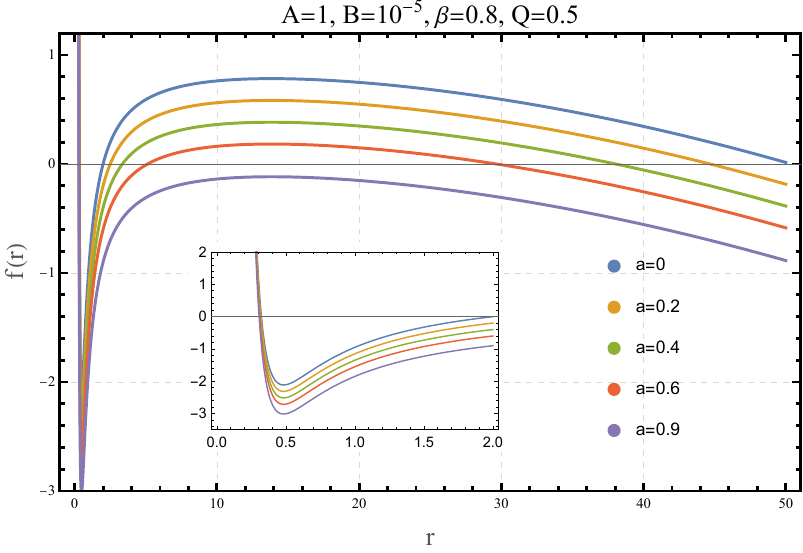}\\
\includegraphics[width=.394\textwidth]{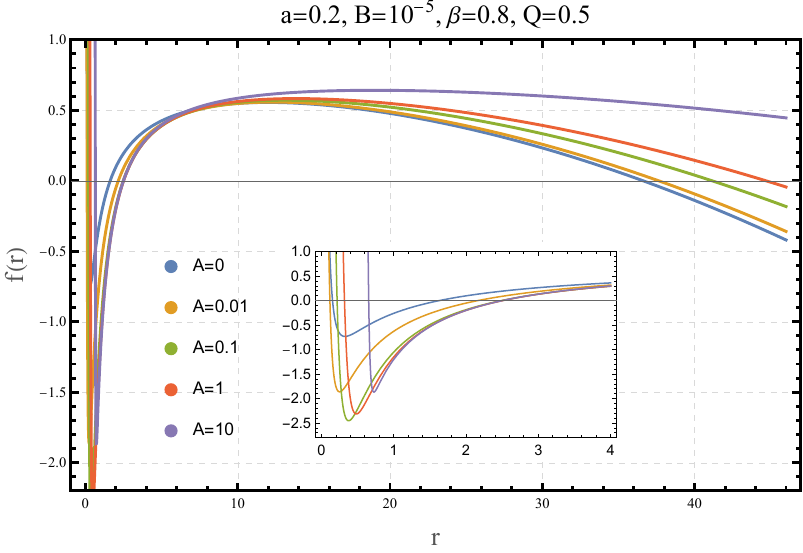}\hspace{0.05\textwidth}
\includegraphics[width=.394\textwidth]{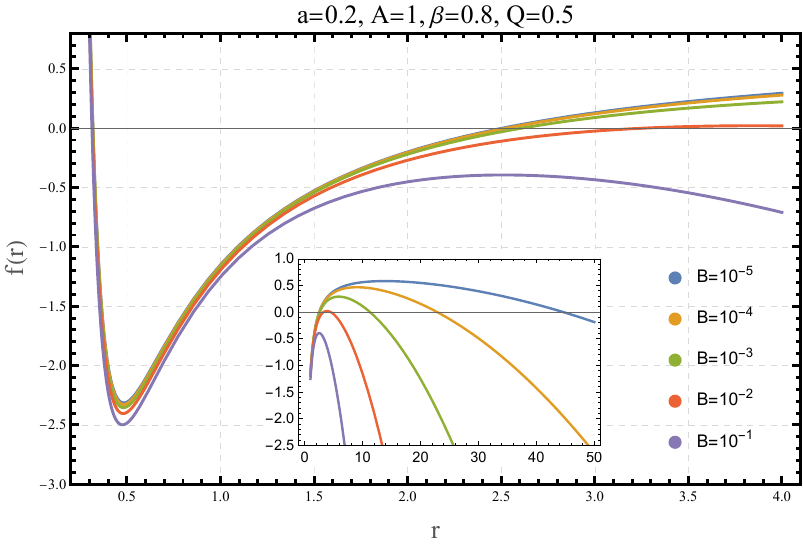}\\
\includegraphics[width=.394\textwidth]{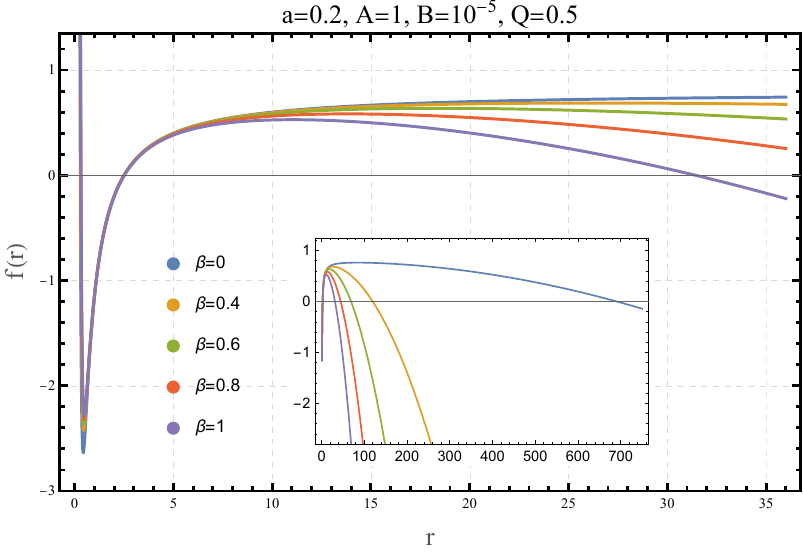}\hspace{0.05\textwidth}
\includegraphics[width=.394\textwidth]{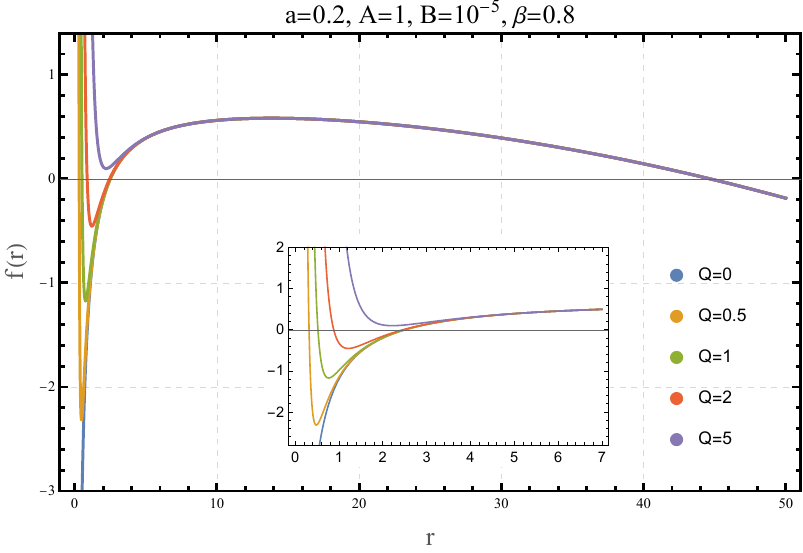}\\ 
\caption{\label{fig:lapsefunction} The lapse function $f(r)$ for diverse parameter configurations. The first panel displays an MCDF-CoS reference case with $a=0.2$, $A=1$, $B=10^{-5}$, $\beta=0.8$, $Q=0.5$, alongside specific limits: MCDF ($a=0$), GCDF ($a=0$, $A=0$, $\beta\neq1$), CDF ($a=0$, $A=0$, $\beta=1$), and Schwarzschild-de Sitter ($a=0$, $Q=0$). Subsequent panels vary individual parameters while maintaining others at reference values.}
\end{figure*}

\section{Black hole immersed in a modified Chaplygin-like dark fluid and a cloud of strings}
\label{sec:blackholebackground}

We consider a static, spherically symmetric spacetime metric describing a black hole immersed in both a modified Chaplygin-like dark fluid (MCDF) and a cloud of strings (CoS). The line element takes the form \cite{Li:2025ftb}
\begin{equation}\label{metric}
    ds^2 = -f(r)dt^2 + \frac{1}{f(r)}dr^2 + r^2 d\Omega^2,
\end{equation}
where $d\Omega^2 = d\theta^2 + \sin^2\theta d\phi^2$ represents the metric on the two-sphere, and $f(r)$ denotes the lapse function:
\begin{equation}
\label{lapsefunction}
    f(r) = 1 - \frac{2M}{r} - a + G(r).
\end{equation}
Here, $a$ parameterizes the CoS contribution, while $G(r)$ encapsulates the MCDF influence, explicitly given by
\begin{equation}\label{mcdfinlapse}
   G(r) = -\frac{r^2}{3}\left(\frac{1+A}{B}\right)^{a_1} ~_2F_1[a_1, a_2; a_3; a_4],
\end{equation}
where $~_2F_1[a_1,a_2;a_3;a_4]$ is the Gauss hypergeometric function with parameters defined as:
\begin{eqnarray}
   a_1 &=& -(1+\beta)^{-1}, \\
   a_2 &=& a_1(1+A)^{-1}, \\
   a_3 &=& 1 + a_2, \\
   a_4 &=& -B^{-1}Q^{-3/a_2} r^{3/a_2}.
\end{eqnarray}
The parameters $A \geq 0$, $B > 0$, and $0 \leq \beta \leq 1$ specify the equation of state of the MCDF \cite{Bento:2002ps, Li:2022csn}:
\begin{equation}
\label{MCDFeos}
    p = A\rho - \frac{B}{\rho^\beta},
\end{equation}
with $Q > 0$ representing the MCDF intensity. 
This model provides a unified description of dark matter and dark energy \cite{Kamenshchik:2001cp, Bilic:2001cg}. In the limit \( A = 0 \), Eq.~\eqref{MCDFeos} reduces to the generalized Chaplygin gas form, \( p = -B/\rho^\beta \) \cite{Bento:2002ps}; further setting \( \beta = 1 \) recovers the original Chaplygin gas form, \( p = -B/\rho \) \cite{Kamenshchik:2001cp}.

The asymptotic behavior of the lapse function, illustrated in Fig.~\ref{fig:lapsefunction} and discussed in Ref.~\cite{Li:2025ftb,Sekhmani:2024rpf}, reveals that as $r \to \infty$:
\begin{equation}\label{asymptoticlapse}
f(r) \to 1 - a - \frac{r^2}{3} \left( \frac{1+A}{B} \right)^{-\frac{1}{1+\beta}}.
\end{equation}
This indicates that the spacetime is asymptotically de Sitter, with an effective cosmological constant shaped by the MCDF parameters: \begin{equation}\label{lambdaMCDF}
\Lambda_{\mathrm{MCDF}} \equiv \left( \frac{1+A}{B} \right)^{-\frac{1}{1+\beta}}.
\end{equation}
However, this metric describes a black hole only within specific parameter ranges that ensure the existence of event horizons. As shown in Fig.~\ref{fig:lapsefunction} and subsequent analysis, black hole solutions require upper bounds on $a$, $B$, and $Q$, and lower bounds on $A$ and $\beta$. Our analysis is accordingly restricted to such physical parameter regions.
For such asymptotically de Sitter black hole spacetimes, both an event horizon $r_h$ and a cosmological horizon $r_c$ exist, corresponding to roots of $f(r)$. Additionally, a Cauchy horizon appears inside the black hole whenever the MCDF intensity $Q$ is nonzero.

When $a=0$, the black hole metric \eqref{metric} describes a black hole immersed solely in an MCDF. In the limit $Q \rightarrow 0$, the MCDF lapse function reduces to
\begin{equation}\label{SdSlapse}
  f(r) \to 1 - \frac{2M}{r} - \frac{r^2}{3} \Lambda_{\mathrm{MCDF}} ,
\end{equation}
which describes a Schwarzschild-de Sitter (S-dS) black hole upon identifying the effective cosmological constant $\Lambda_{\rm MCDF}$ [Eq.~\eqref{lambdaMCDF}]
with the standard cosmological constant $\Lambda$.
This connection highlights the role of the MCDF in driving cosmic acceleration. The specific cases of \( (A=0,~ \beta \neq 1) \) and \( (A=0,~ \beta = 1) \) yield the GCDF and CDF black hole metrics, respectively. It is noteworthy that while \( \Lambda_{\mathrm{MCDF}} \) fixes the large-\( r \) asymptotic structure, the parameter \( A \) predominantly influences the near-horizon geometry and the precise location of the event horizon, especially for small values of \( A \).

In subsequent analysis, we set $M=1$ for computational convenience.

\section{Shadow and optical appearance}
\label{sec:shadow_appearance}

\subsection{Critical photon orbits and black hole shadow}

The propagation of null geodesics with energy $E$ and angular momentum $L$ is governed by \cite{Li:2025ftb} % Please add reference
\begin{eqnarray}
    \label{teq}
    \dot t &=& \frac{E}{f(r)}, \\
    \label{phieq}
    \dot\phi &=& \frac{L}{r^2}, \\
    \label{radialeq}
    \dot r^2 + \mathcal{V}(r) &=& E^2,
\end{eqnarray}
where dots denote derivatives with respect to the affine parameter $s$, and
\begin{equation}\label{geodesicpotential}
    \mathcal{V}(r) = \frac{L^2}{r^2} f(r)
\end{equation}
is the effective potential for radial motion. The explicit energy dependence can be scaled out, reducing the photon trajectory description to a single parameter: the impact parameter $b \equiv L/E$.

The photon sphere radius, comprising unstable spherical photon orbits, is determined by
\begin{equation}\label{rpheq}
    f(r_\text{ps}) - \frac{1}{2} r_\text{ps} f'(r_\text{ps}) = 0,
\end{equation}
where primes indicate radial derivatives. The corresponding critical impact parameter and angular velocity are:
\begin{eqnarray}
\label{criticalimpactparameter}
    b_\text{ps} &=& \frac{r_\text{ps}}{\sqrt{f(r_\text{ps})}}, \\
 \label{angularvelocity}
    \Omega_\text{ps} &=& \frac{\sqrt{f(r_\text{ps})}}{r_\text{ps}},
\end{eqnarray}
while the Lyapunov exponent, characterizing orbital instability, reads:
\begin{equation}
\label{Lyapunov}
    \lambda = \frac{1}{\sqrt{2}}\sqrt{\frac{f(r)}{r^2}\Big(2f(r)-r^2  f^{\prime\prime}(r)\Big)}~\Bigg|_{r=r_\text{ps}}.
\end{equation}
%\begin{equation}
 %   \lambda = \frac{1}{\sqrt{2}} \sqrt{ \frac{r_\text{ps}^2}{f(r_\text{ps})} \left( \frac{d^2}{dr_*^2} \frac{f}{r^2} \right)_{r=r_\text{ps}} },
%\end{equation}
%with the tortoise coordinate $r_*$ defined as   

For a distant observer at $r_O$, the shadow radius becomes:
\begin{equation}\label{shadowradius}
    R_s = b_\text{ps}\sqrt{f(r_O)}.
\end{equation}

\subsection{Optical appearance under spherical accretion}

While previous studies examined MCDF-CoS black holes with thin accretion disks \cite{Li:2025ftb}, we focus here on spherical accretion models. The specific intensity observed at $r = r_O$ is given by \cite{Jaroszynski:1997bw, Bambi:2013nla}:
\begin{equation} \label{intensity}
I = \int_\gamma g^3 j(\nu_{\rm e}) dl_{\rm p},
\end{equation}
where $g = \nu_{\rm o}/\nu_{\rm e}$ is the redshift factor, $j(\nu_{\rm e})$ the emissivity per unit volume in the emitter's frame, $dl_{\rm p}$ the infinitesimal proper length, and $\gamma$ the light ray trajectory.

\subsubsection{Static accretion model}

For static spherical accretion, $g = [f(r)/f(r_{\rm O})]^{1/2}$. Assuming monochromatic emission at frequency $\nu_s$, the specific emissivity follows:
\begin{equation}
j(\nu_{\rm e}) \propto \delta(\nu_e - \nu_s) r^{-m}, \label{profile}
\end{equation}
where we adopt $m=2$ following established conventions \cite{Bambi:2013nla}, though $m=4$ \cite{Kocherlakota:2022jnz} and $m=6$ \cite{Bauer:2021atk} are also physically motivated. The proper length element is:
\begin{eqnarray}
dl_{\rm p} &=& \sqrt{ f(r)^{-1} dr^2 + r^2 d\phi^2 } \nonumber \\
  &=& \sqrt{ f(r)^{-1} + r^2 \left( \frac{d\phi}{dr} \right)^2 } dr, \label{dl}
\end{eqnarray}
with $d\phi/dr$ derivable from Eqs.~\eqref{phieq} and \eqref{radialeq}. The observed specific intensity thus becomes:
\begin{equation}\label{staticintensity}
I_{\rm{obs}} = \int_\gamma \left[ \frac{f(r)}{f(r_{\rm O})} \right]^{3/2} r^{-2} \sqrt{ f(r)^{-1} + r^2 \left( \frac{d\phi}{dr} \right)^2 } dr.
\end{equation}

\subsubsection{Infalling accretion model}

For infalling accretion, the redshift factor incorporates flow velocity effects:
\begin{equation}
g = \frac{k_\beta u^\beta_{\rm O}}{k_\gamma u^\gamma_{\rm e}}, \label{redf}
\end{equation}
where $k^\mu = \dot{x}_\mu$ is the photon four-velocity, $u^\mu_{\rm O} = (f(r_{\rm O})^{-1/2}, 0, 0, 0)$ the static observer's four-velocity, and 
\begin{eqnarray}
u^t_{\rm e} = \frac{1}{f(r)}, \quad
u^r_{\rm e} = - \sqrt{ 1 - f(r) }, \quad
u^\theta_{\rm e} = u^\phi_{\rm e} = 0
\end{eqnarray}
the accretion flow four-velocity. For null geodesics with affine parameter $s \to s/|L|$, $k_t = 1/b$ remains constant, while $k_r$ follows from $k_\alpha k^\alpha = 0$:
\begin{equation}
\frac{k_r}{k_t} = \pm \frac{1}{f(r)} \sqrt{ 1 - \frac{b^2 f(r)}{r^2} }, \label{krkt}
\end{equation}
with signs distinguishing between incoming ($-$) and outgoing ($+$) photons. This yields the simplified redshift factor:
\begin{equation}
g = \frac{1}{u^t_{\rm e} + (k_r/k_t) u^r_{\rm e}}, \label{sredf}
\end{equation}
which differs fundamentally from the static case.

The proper distance becomes 
\begin{equation}
dl_{\mathrm {p}} = k_\mu u^\mu_{\mathrm {e}} ds = \frac{k_t}{g |k_r|} dr,
\end{equation}
leading to the observed intensity:
\begin{equation}
I_{\rm{obs}} \propto \int_\gamma \frac{g^3 k_t dr}{r^2 |k_r|}. \label{infallingintensity}
\end{equation}
We employ $f(r_{\rm O})^{3/2} I_{\rm obs}(b)$ for numerical analysis to remove the observer-position dependence. The absolute value $|k_r|$ accounts for photon direction reversals along the trajectory.

\begin{figure*}[h]
\centering % \begin{center}/\end{center} takes some additional vertical space
\includegraphics[width=.374\textwidth]{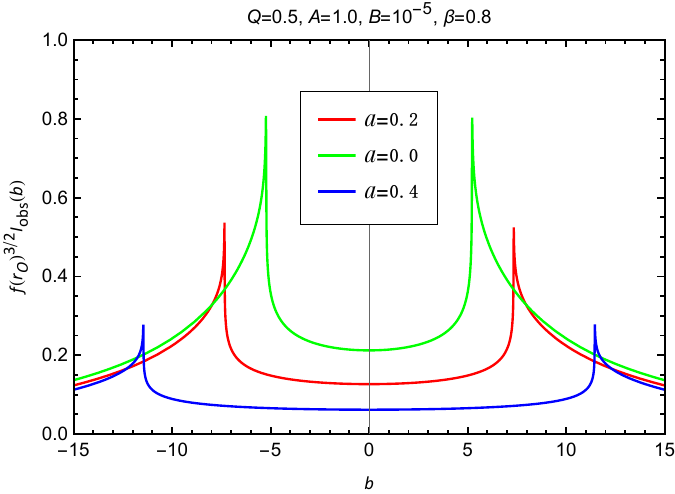}\hspace{0.05\textwidth}
\includegraphics[width=.305\textwidth]{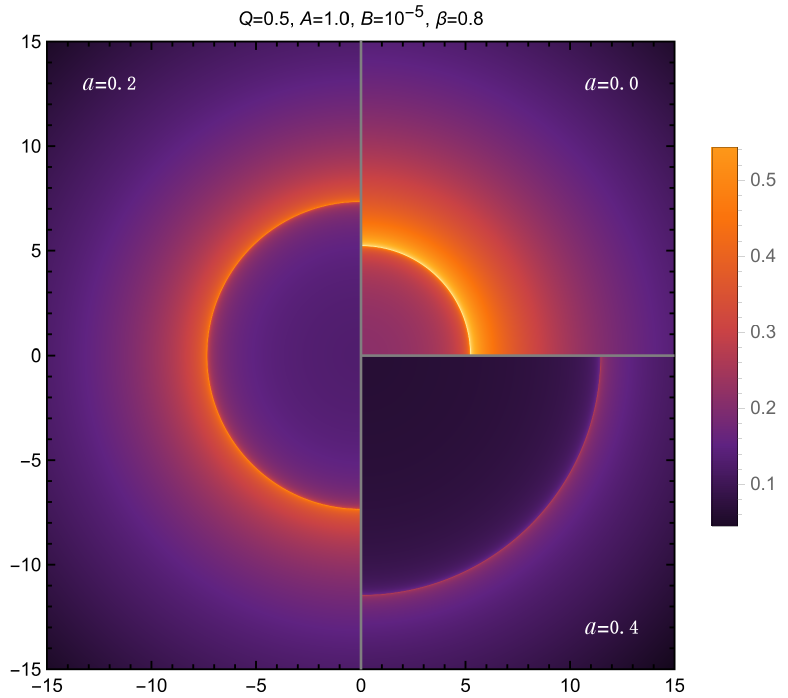}\\
\includegraphics[width=.374\textwidth]{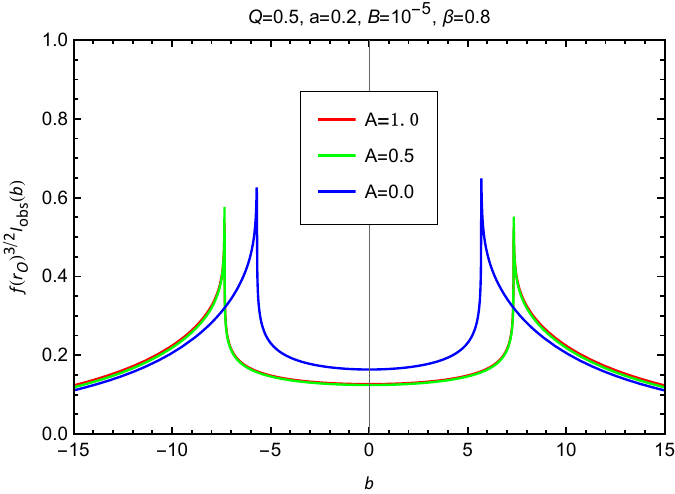}\hspace{0.05\textwidth}
\includegraphics[width=.305\textwidth]{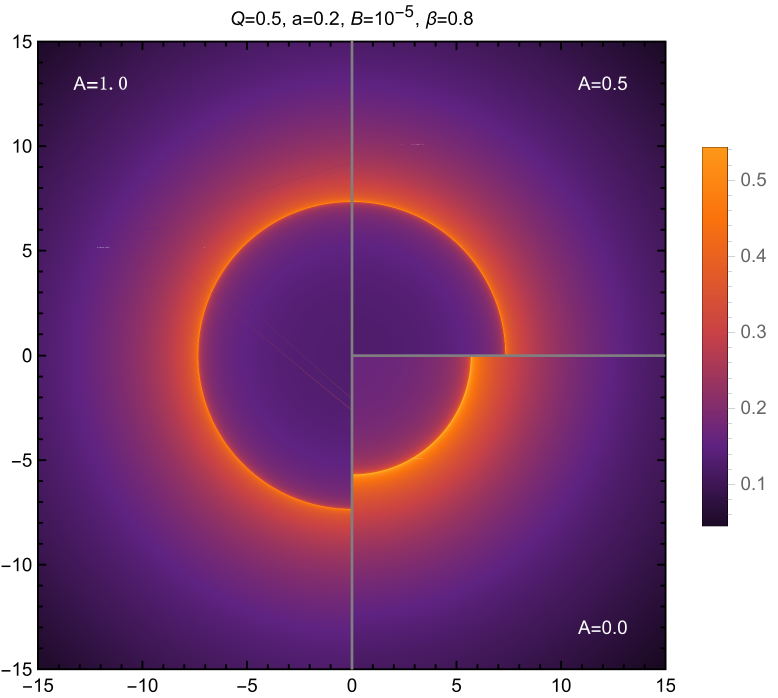}\\
\includegraphics[width=.374\textwidth]{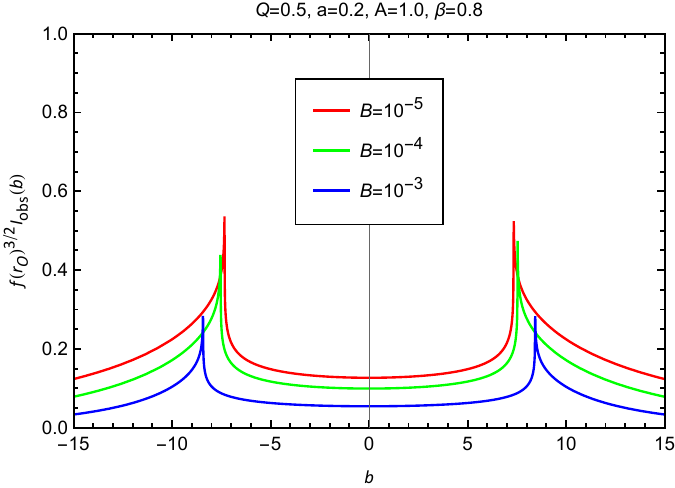}\hspace{0.05\textwidth}
\includegraphics[width=.305\textwidth]{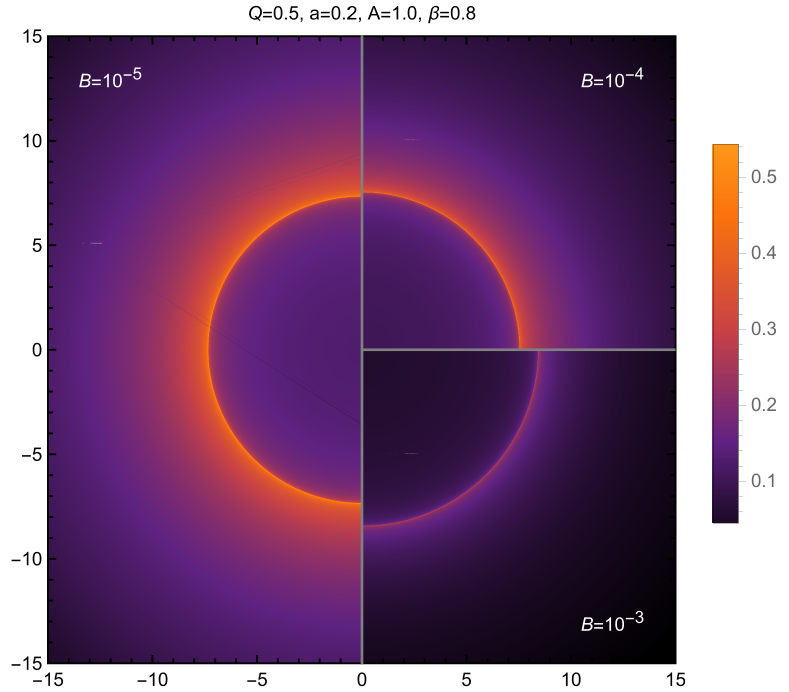}\\
\includegraphics[width=.374\textwidth]{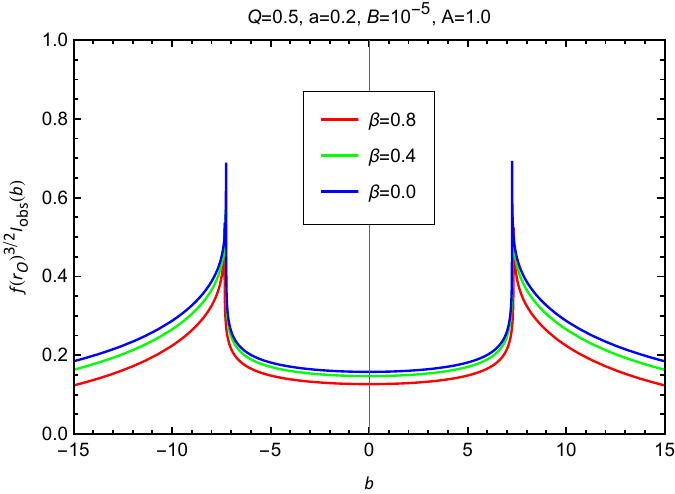}\hspace{0.05\textwidth}
\includegraphics[width=.305\textwidth]{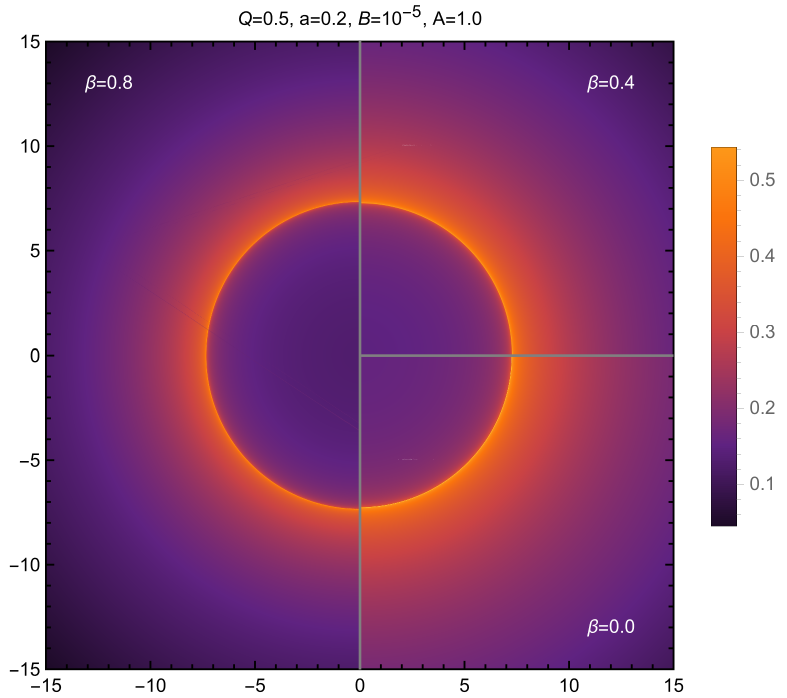}\\
\includegraphics[width=.374\textwidth]{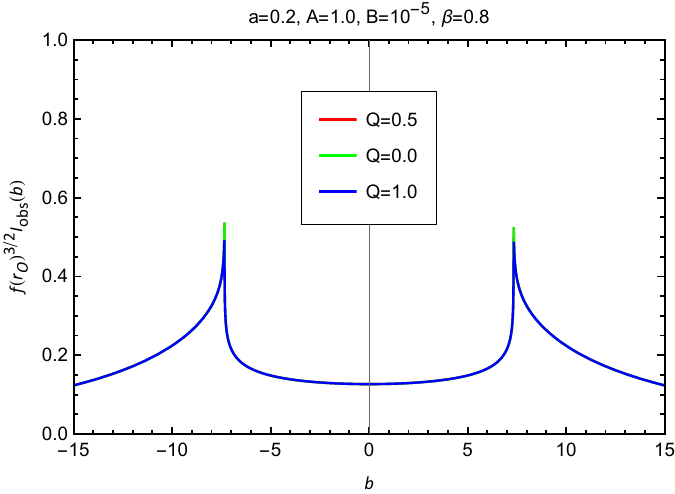}\hspace{0.05\textwidth}
\includegraphics[width=.305\textwidth]{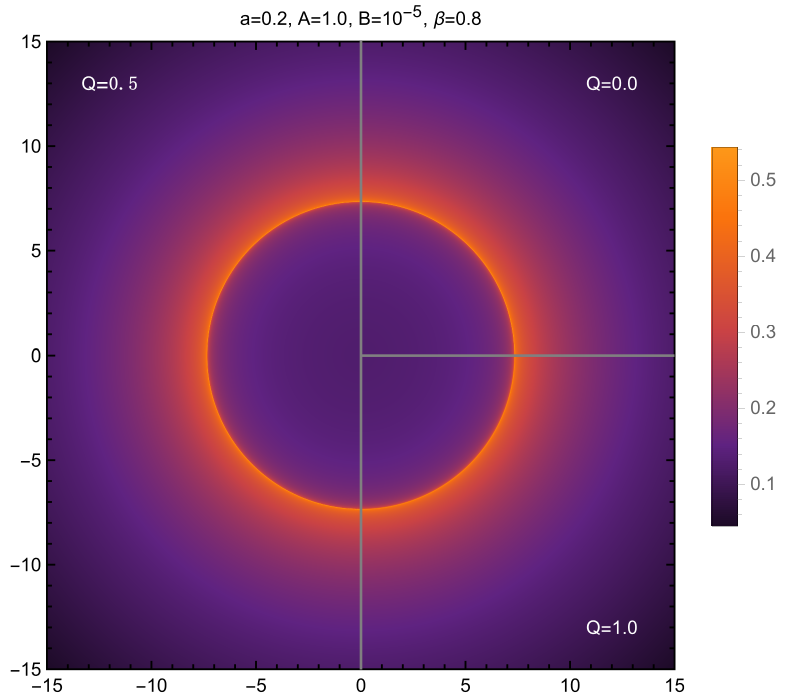}\\ 
\caption{\label{fig:staticimages}
Profiles of the specific intensity $I_{\rm{obs}}(b)$ (left panels) and corresponding images (right panels) for static spherical accretion, viewed face-on by an observer near the pseudo-cosmological horizon. }
\end{figure*}

\begin{figure*}[]
\centering % \begin{center}/\end{center} takes some additional vertical space
\includegraphics[width=.374\textwidth]{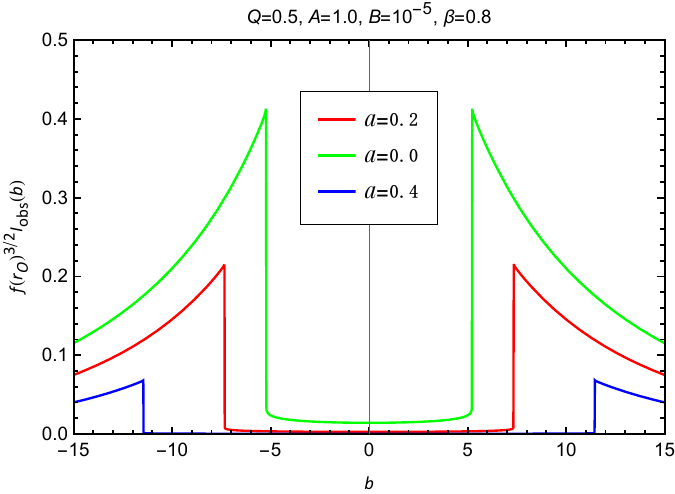}\hspace{0.05\textwidth}
\includegraphics[width=.305\textwidth]{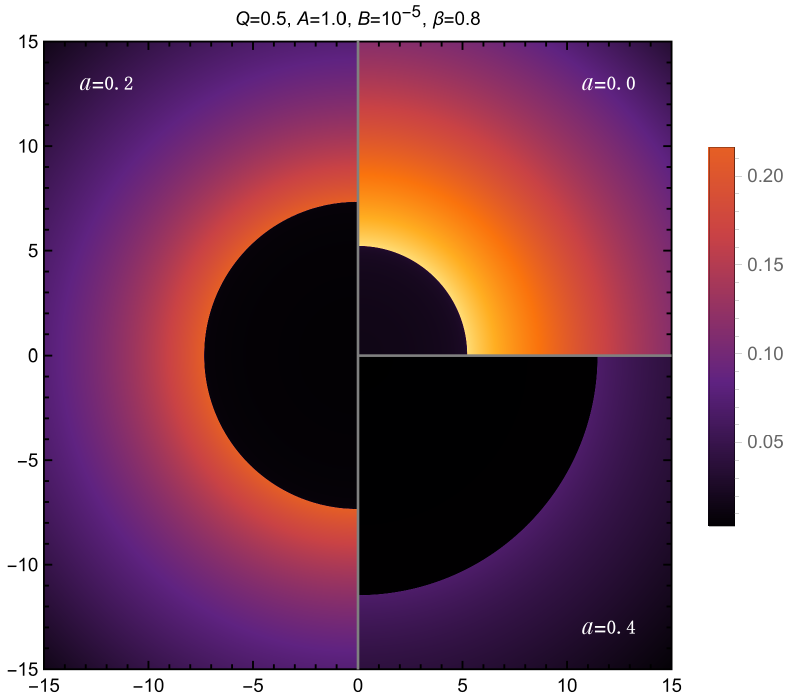}\\
\includegraphics[width=.374\textwidth]{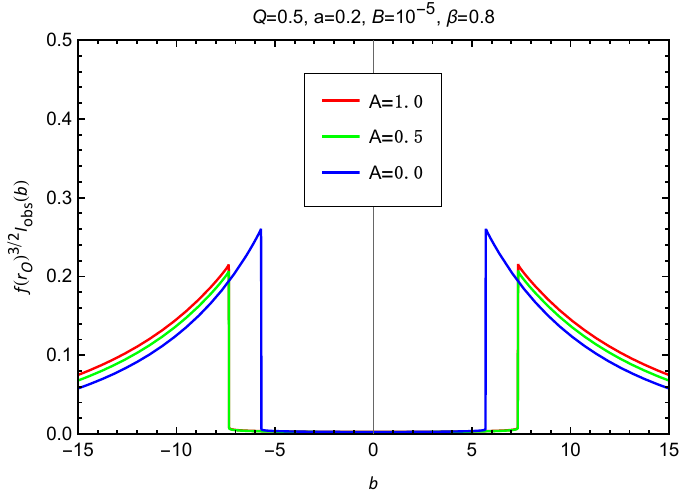}\hspace{0.05\textwidth}
\includegraphics[width=.305\textwidth]{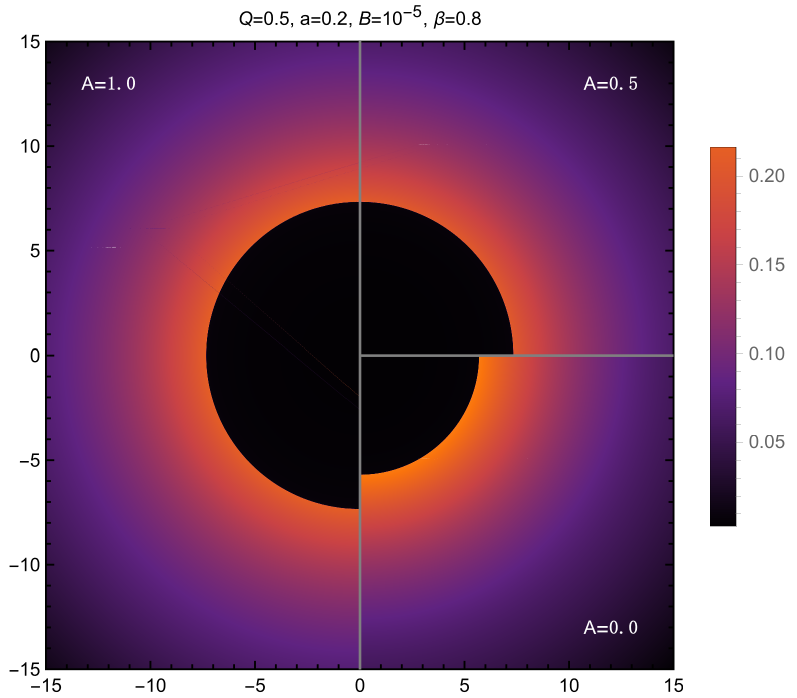}\\
\includegraphics[width=.374\textwidth]{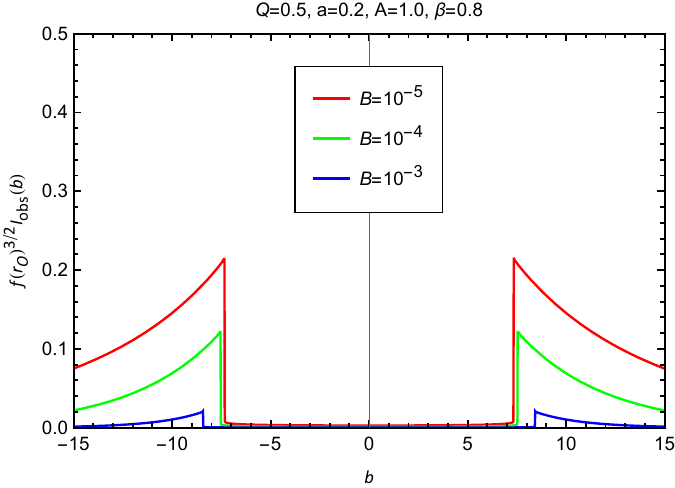}\hspace{0.05\textwidth}
\includegraphics[width=.305\textwidth]{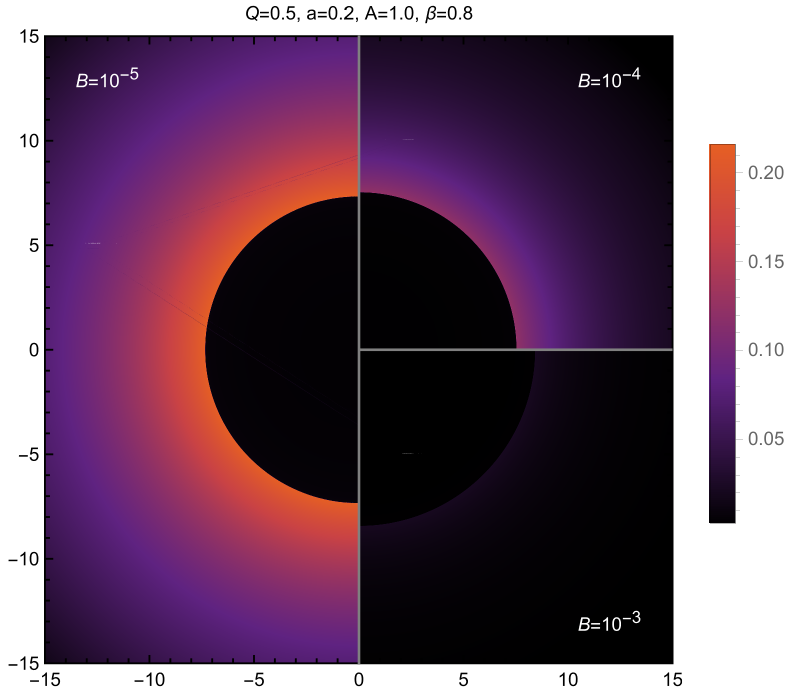}\\
\includegraphics[width=.374\textwidth]{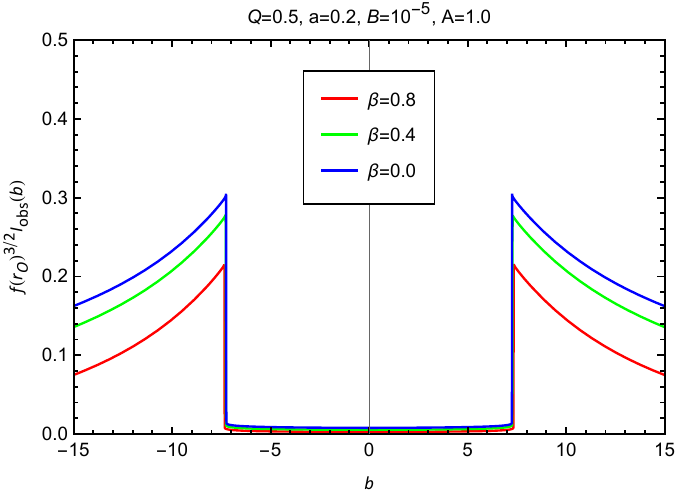}\hspace{0.05\textwidth}
\includegraphics[width=.305\textwidth]{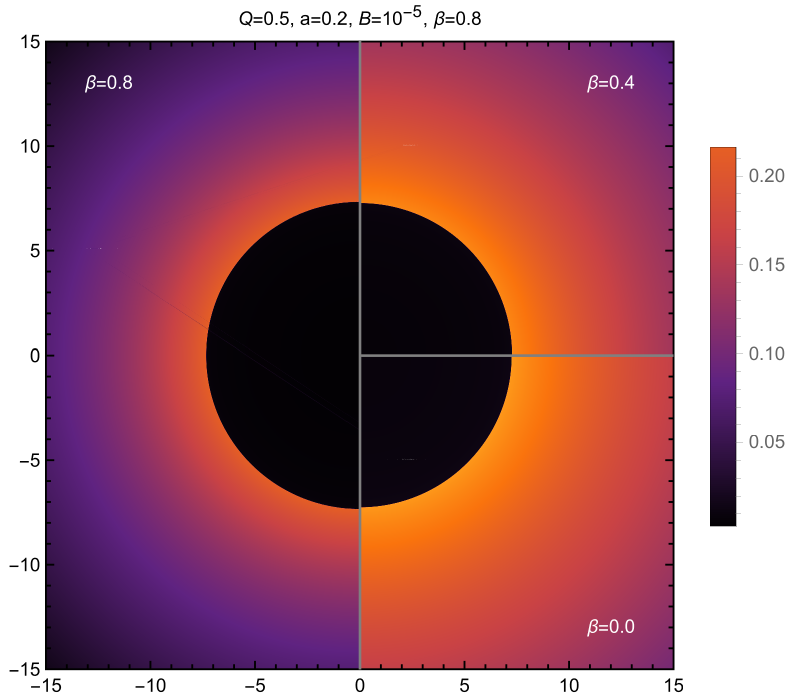}\\
\includegraphics[width=.374\textwidth]{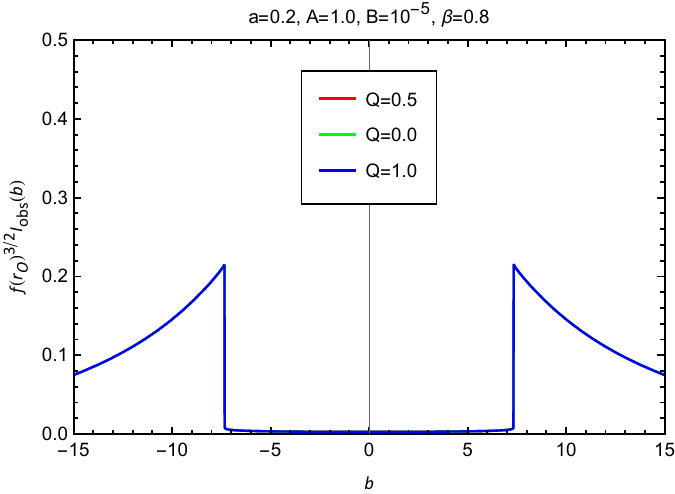}\hspace{0.05\textwidth}
\includegraphics[width=.305\textwidth]{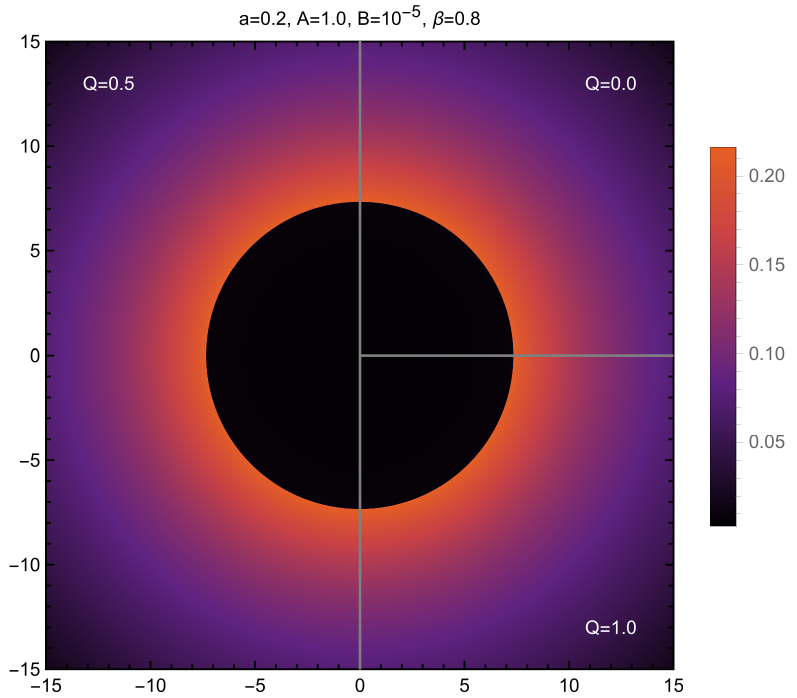}\\ 
\caption{\label{fig:infallingimages} 
Profiles of the specific intensity $I_{\rm{obs}}(b)$ (left panels) and corresponding images (right panels) for infalling spherical accretion, viewed face-on by an observer near the pseudo-cosmological horizon.}
\end{figure*}

\subsubsection{Optical appearance analysis}
\label{subsec:optical_appearance}

Our analysis of the MCDF-CoS black hole's optical appearance, based on numerical integration of specific intensities for static \eqref{staticintensity} and infalling \eqref{infallingintensity} accretion models, reveals characteristic features. The observed intensity $f(r_{\rm O})^{3/2}I_{\rm obs}(b)$, plotted against impact parameter $b$ in Figs.~\ref{fig:staticimages} and \ref{fig:infallingimages}, exhibits a universal profile: intensity rises sharply with $b$, peaks at the photon sphere $b = b_{\rm ps}$, then declines rapidly. This defines the black hole shadow—a central dark region ($b < b_{\rm ps}$) where most photons are absorbed by the event horizon. The intensity maximum at $b_{\rm ps}$ results from photons undergoing multiple unstable orbits, accumulating enhanced path lengths through the emitting medium. For $b > b_{\rm ps}$, intensity derives from refracted rays and diminishes with increasing $b$.

Despite sharing the identical shadow radius $b_{\rm ps}\sqrt{f(r_O)}$ (a geometric invariant), the two accretion models produce distinct observational signatures. Static accretion yields a sharper, narrower, brighter emission ring, while infalling accretion generates a broader, fainter ring with a darker shadow interior. This suppression in the infalling case manifests Doppler beaming: radially infalling matter redshifts forward-emitted radiation, reducing detected flux at infinity.

Parameter variations systematically affect the shadow and emission ring. The CoS parameter $a$ exerts the strongest influence: increasing $a$ significantly enlarges the shadow radius while reducing intensity. MCDF parameters produce more nuanced effects: $Q$ has negligible impact; decreasing $\beta$ slightly enhances intensity without affecting the shadow radius; only extreme $A$ and $B$ values noticeably alter the images. Specifically, increasing $A$ enlarges the shadow and reduces intensity (both effects saturating at large $A$), while increasing $B$ reduces both shadow size and intensity. These consistent trends across accretion models confirm that spacetime geometry primarily determines shadow properties, while accretion dynamics shape the bright ring characteristics.

\section{Quasinormal modes}
\label{sec:qnm}

We investigate the propagation of a massless scalar field on the fixed background of the MCDF-CoS black hole spacetime, described by the metric \eqref{metric}. The dynamics of the scalar perturbation are governed by the covariant Klein-Gordon equation \cite{Berti:2009kk,Konoplya:2011qq}:
\begin{equation}\label{KGeq}
\nabla_\mu \nabla^\mu \Phi = \frac{1}{\sqrt{-g}} \partial_\mu \left( \sqrt{-g} g^{\mu\nu} \partial_\nu \Phi \right) = 0.
\end{equation}
Owing to the spherical symmetry of the background, the field equation admits separation of variables via the ansatz:
\begin{equation}\label{ansatz}
    \Phi(t, r, \theta, \phi) = \frac{\Psi(r)}{r} Y_{lm}(\theta, \phi) e^{-i\omega t},
\end{equation}
where $Y_{lm}(\theta, \phi)$ are the standard spherical harmonics, and $\omega$ is the complex quasinormal mode (QNM) frequency. The real part, $\omega_R$, corresponds to the oscillation frequency, while the imaginary part, $\omega_I$, determines the damping rate. Substituting this ansatz into Eq.~\eqref{KGeq} leads to a Schrödinger-like wave equation for the radial function $\Psi(r)$:
\begin{equation}\label{waveeq}
    \frac{d^2\Psi}{dr_*^2} + \left[ \omega^2 - V(r) \right] \Psi = 0,
\end{equation}
governed by the effective potential
\begin{equation}
\label{potential}
    V(r) = f(r) \left[ \frac{l(l+1)}{r^2} + \frac{1}{r} \frac{df(r)}{dr} \right].
\end{equation}
Here, the tortoise coordinate $r_*$ is defined by %$dr_* = dr/f(r)$,
\begin{equation}
\label{tortoisecoor}
dr_* = \frac{dr}{f(r)},
\end{equation}
which maps the event horizon $r_h$ and the cosmological horizon $r_c$ to $r_* \to -\infty$ and $r_* \to +\infty$, respectively, ensuring that $V(r) \to 0$ at both boundaries.

The QNM spectrum consists of discrete complex eigenvalues $\{\omega_n\}$, which are determined by imposing purely outgoing wave boundary conditions at both horizons:
\begin{eqnarray}
   \Psi(r_*) \sim e^{-i\omega r_*} \quad \text{as} \quad r_* \to -\infty \quad (\text{near } r_h), \\
   \Psi(r_*) \sim e^{+i\omega r_*} \quad \text{as} \quad r_* \to +\infty \quad (\text{near } r_c).
\end{eqnarray}
These boundary conditions render the eigenvalue problem non-self-adjoint, leading to a discrete set of complex frequencies $\omega = \omega_R + i \omega_I$ that encode the intrinsic, model-dependent ringing of the spacetime.

\begin{figure*}[]
\centering % \begin{center}/\end{center} takes some additional vertical space
\includegraphics[width=.374\textwidth]{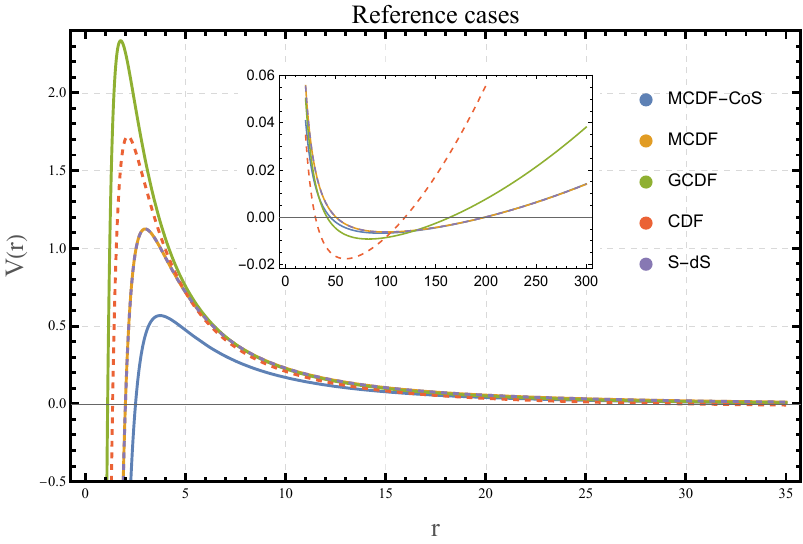}\hspace{0.05\textwidth}
\includegraphics[width=.25\textwidth]{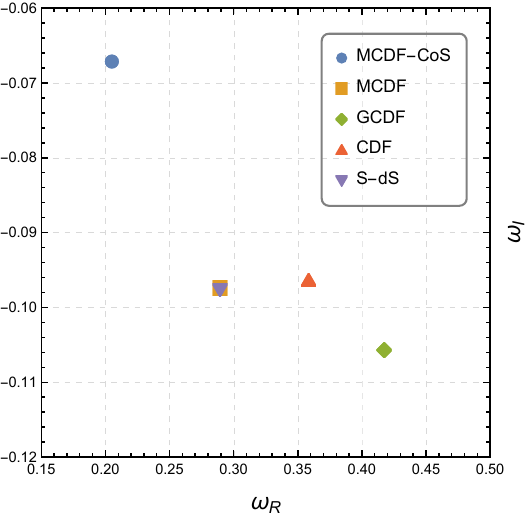}\\
\caption{\label{fig:referencemodes}  
Effective potentials (left) and corresponding fundamental ($n=0$, $l=1$) QNM frequencies (right) for the reference models. Model parameters are detailed in Table~\ref{tab:referencemodes}.}
\end{figure*}

\begin{table*}[]
    \centering
    \caption{Comparison of fundamental ($n=0$, $l=1$) QNM frequencies for different black hole models. The parameters for the reference MCDF-CoS case are: $a=0.2$, $A=1$, $B=10^{-5}$, $\beta=0.8$, $Q=0.5$. Other cases represent specific limits of this general configuration. $\Delta_R\%$ and $\Delta_I\%$ denote the relative errors for real and imaginary parts, calculated as $|\omega_{\mathrm{WKB}} - \omega_{\mathrm{P-T}}|/|\omega_{\mathrm{WKB}}| \times 100\%$.}
    \label{tab:referencemodes}
    \scriptsize
    \renewcommand{\arraystretch}{1.6}
    \setlength{\tabcolsep}{6pt}
    \begin{tabular}{@{}l c c c c c c c c c@{}}
        \toprule
        \textbf{Model} & $a$ & $A$ & $B$ & $\beta$ & $Q$ & $\omega_{\mathrm{WKB}}$ & $\omega_{\mathrm{P-T}}$ & $\Delta_R\%$ & $\Delta_I\%$ \\
        \midrule
        MCDF-CoS & 0.2 & 1 & $10^{-5}$ & 0.8 & 0.5 & 0.203417 - 0.062029 $\mi$ & 0.207452 - 0.063452 $\mi$ & 1.96 & 2.29 \\
        MCDF      & 0   & 1 & $10^{-5}$ & 0.8 & 0.5 & 0.289203 - 0.097471 $\mi$ & 0.296591 - 0.100235 $\mi$ & 2.56 & 2.84 \\
        GCDF      & 0   & 0 & $10^{-5}$ & 0.8 & 0.5 & 0.417415 - 0.105700 $\mi$ & 0.422366 - 0.107314 $\mi$ & 1.19 & 1.53 \\
        CDF       & 0   & 0 & $10^{-5}$ & 1   & 0.5 & 0.358448 - 0.096342 $\mi$ & 0.363282 - 0.098002 $\mi$ & 1.35 & 1.72 \\
        S-dS      & 0   & 1 & $10^{-5}$ & 0.8 & 0   & 0.289352 - 0.097672 $\mi$ & 0.296573 - 0.100285 $\mi$ & 2.50 & 2.68 \\
        \bottomrule
    \end{tabular}
\end{table*}

\begin{table*}[]
    \centering
    \scriptsize
    \renewcommand{\arraystretch}{1.4}
    \setlength{\tabcolsep}{6pt}
    \caption{Dependence of the fundamental ($n=0$, $l=1$) QNM frequencies on various metric parameters. Each section varies one parameter while keeping others fixed at the reference values: $a=0.2$, $A=1$, $B=10^{-5}$, $\beta=0.8$, $Q=0.5$ (unless otherwise specified in the Fixed Parameters column). $\Delta_R\%$ and $\Delta_I\%$ denote the relative errors for real and imaginary parts, calculated as $|\omega_{\mathrm{WKB}} - \omega_{\mathrm{P-T}}|/|\omega_{\mathrm{WKB}}| \times 100\%$.}
    \label{tab:qnms_metricdependence}
    \begin{tabular}{c c c c c c}
        \toprule
        \textbf{Fixed Parameters} & \textbf{Variable Parameter} & $\omega_{\mathrm{WKB}}$ & $\omega_{\mathrm{P-T}}$ & $\Delta_R\%$ & $\Delta_I\%$ \\
        \midrule
        \multirow{3}{*}{\makecell{$Q=0.5$ \\ $A=1.0$ \\ $B=10^{-5}$ \\ $\beta=0.8$}} 
        & $a=0.0$ & 0.289203 - 0.097471 $\mi$ & 0.296591 - 0.100235 $\mi$ & 2.56 & 2.84 \\
        \addlinespace
        & $a=0.2$ & 0.203417 - 0.062029 $\mi$ & 0.207452 - 0.063452 $\mi$ & 1.96 & 2.29 \\
        \addlinespace
        & $a=0.4$ & 0.128447 - 0.034367 $\mi$ & 0.130274 - 0.034983 $\mi$ & 1.42 & 1.79 \\
        \midrule
        \multirow{3}{*}{\makecell{$Q=0.5$ \\ $B=10^{-5}$ \\ $\beta=0.8$ \\ $a=0.2$}} 
        & $A=0$   & 0.262068 - 0.062967 $\mi$ & 0.264851 - 0.063863 $\mi$ & 1.06 & 1.42 \\
        \addlinespace
        & $A=0.5$ & 0.203224 - 0.061391 $\mi$ & 0.207297 - 0.062959 $\mi$ & 2.00 & 2.55 \\
        \addlinespace
        & $A=1.0$ & 0.203417 - 0.062029 $\mi$ & 0.207452 - 0.063452 $\mi$ & 1.96 & 2.29 \\
        \midrule
        \multirow{3}{*}{\makecell{$Q=0.5$ \\ $A=1$ \\ $\beta=0.8$ \\ $a=0.2$}} 
        & $B=10^{-5}$ & 0.203417 - 0.062029 $\mi$ & 0.207452 - 0.063452 $\mi$ & 1.96 & 2.29 \\
        \addlinespace
        & $B=10^{-4}$ & 0.197131 - 0.060818 $\mi$ & 0.200639 - 0.062128 $\mi$ & 1.78 & 2.15 \\
        \addlinespace
        & $B=10^{-3}$ & 0.173269 - 0.055551 $\mi$ & 0.175396 - 0.056418 $\mi$ & 1.23 & 1.56 \\
        \midrule
        \multirow{3}{*}{\makecell{$Q=0.5$ \\ $A=1$ \\ $B=10^{-5}$ \\ $a=0.2$}} 
        & $\beta=0$   & 0.205887 - 0.062428 $\mi$ & 0.210128 - 0.063886 $\mi$ & 2.06 & 2.34 \\
        \addlinespace
        & $\beta=0.4$ & 0.205472 - 0.062347 $\mi$ & 0.209709 - 0.063828 $\mi$ & 2.06 & 2.37 \\
        \addlinespace
        & $\beta=0.8$ & 0.203417 - 0.062029 $\mi$ & 0.207452 - 0.063452 $\mi$ & 1.96 & 2.29 \\
        \midrule
        \multirow{3}{*}{\makecell{$A=1$ \\ $B=10^{-5}$ \\ $\beta=0.8$ \\ $a=0.2$}} 
        & $Q=0$   & 0.203428 - 0.062043 $\mi$ & 0.207450 - 0.063458 $\mi$ & 1.96 & 2.28 \\
        \addlinespace
        & $Q=0.5$ & 0.203417 - 0.062029 $\mi$ & 0.207452 - 0.063452 $\mi$ & 1.96 & 2.29 \\
        \addlinespace
        & $Q=1$   & 0.203276 - 0.061806 $\mi$ & 0.207480 - 0.063383 $\mi$ & 2.07 & 2.55 \\
        \bottomrule
    \end{tabular}
\end{table*}

\begin{figure*}[] 
\centering % \begin{center}/\end{center} takes some additional vertical space
\includegraphics[width=.375\textwidth]{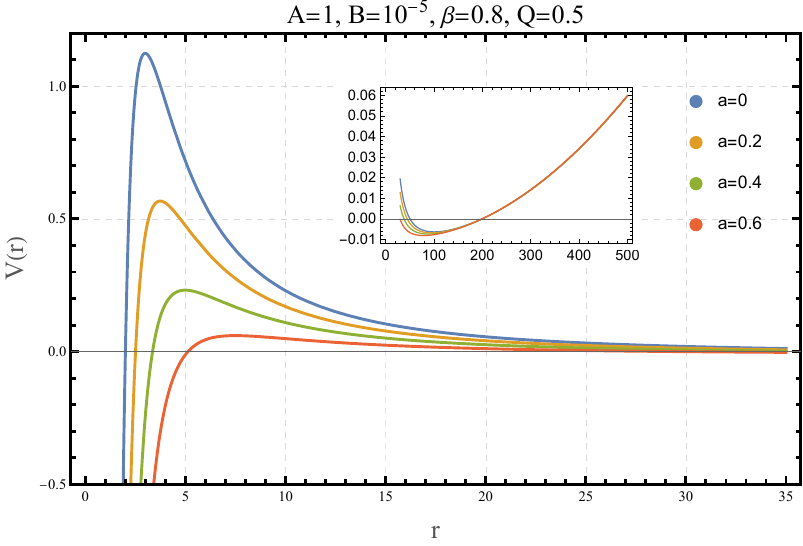}\hspace{0.03\textwidth}
\includegraphics[width=.23\textwidth]{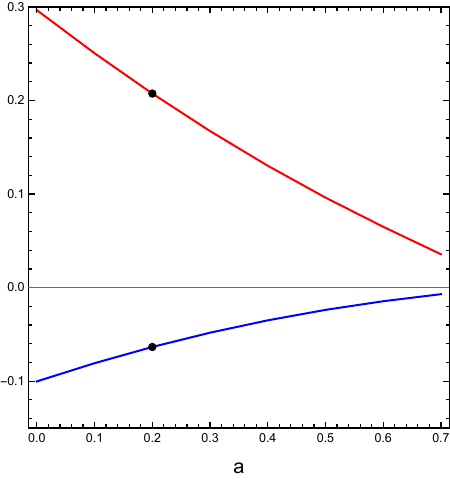}\hspace{0.03\textwidth}\includegraphics[width=.255\textwidth]{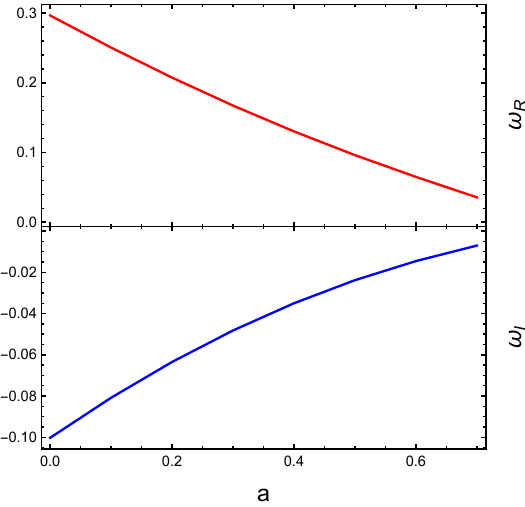}\\
\includegraphics[width=.375\textwidth]{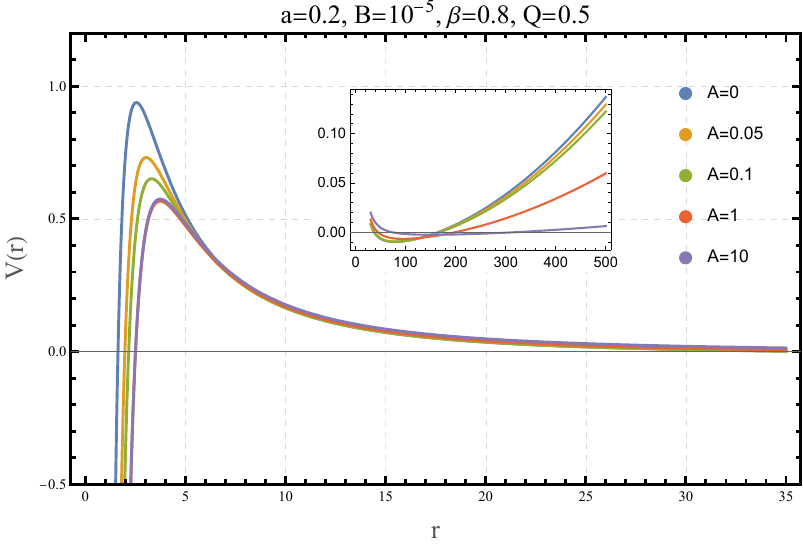}\hspace{0.03\textwidth}
\includegraphics[width=.23\textwidth]{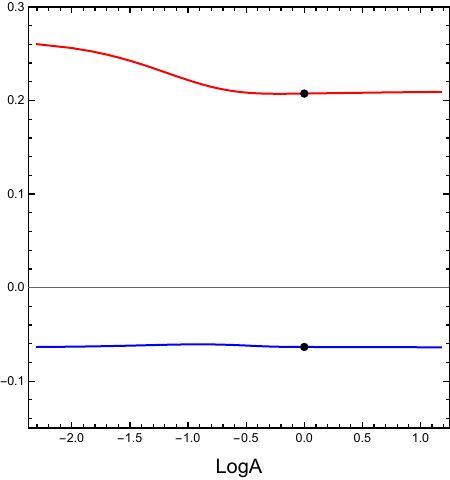}\hspace{0.03\textwidth}\includegraphics[width=.255\textwidth]{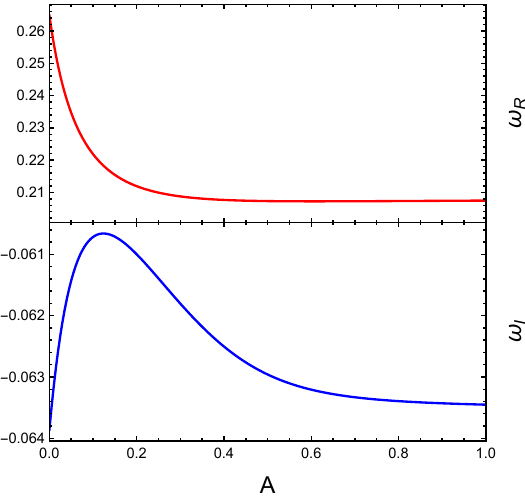}\\
\includegraphics[width=.375\textwidth]{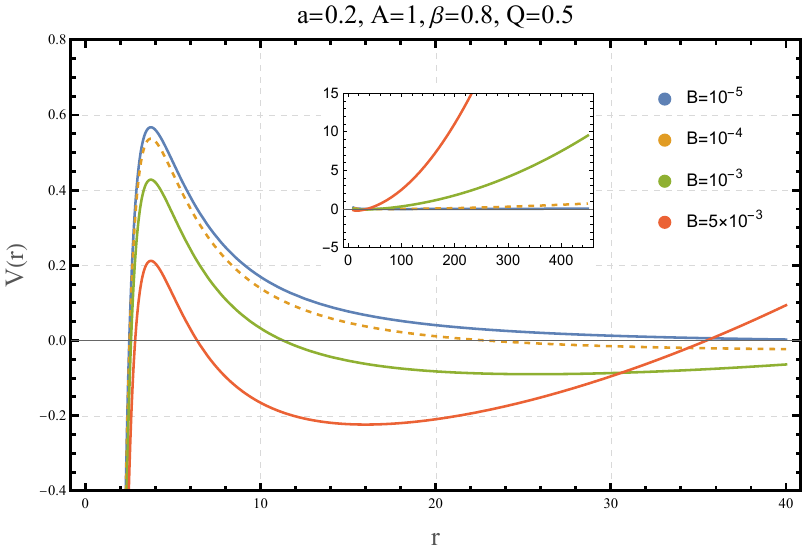}\hspace{0.03\textwidth}
\includegraphics[width=.23\textwidth]{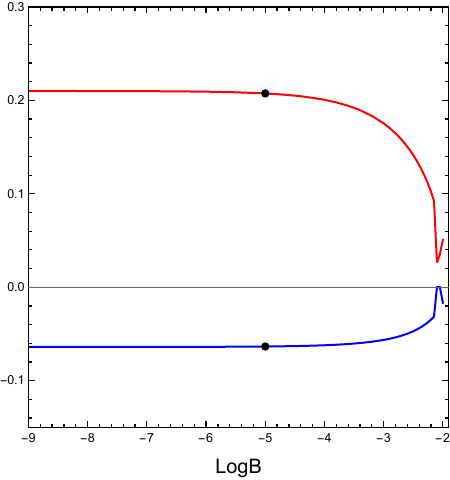}\hspace{0.03\textwidth}\includegraphics[width=.255\textwidth]{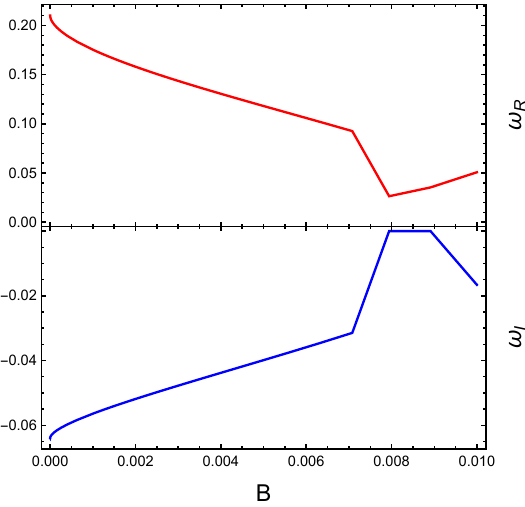}\\
\includegraphics[width=.375\textwidth]{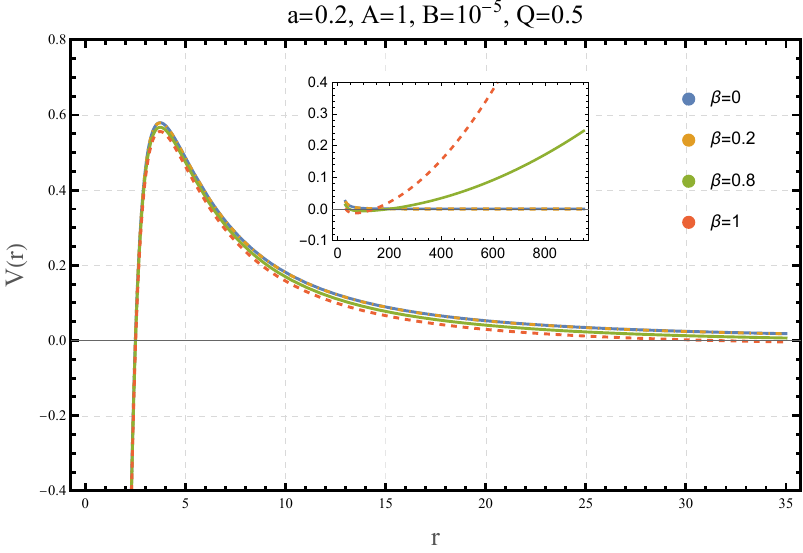}\hspace{0.03\textwidth}
\includegraphics[width=.23\textwidth]{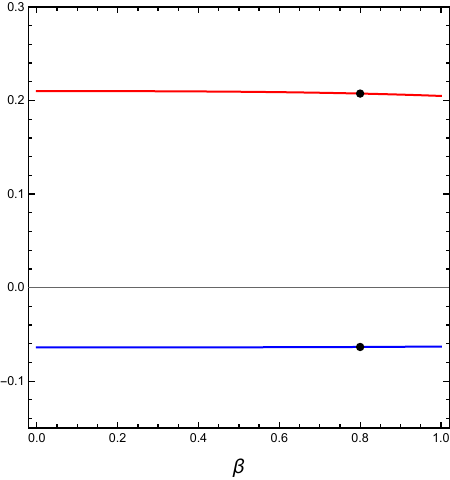}\hspace{0.03\textwidth}\includegraphics[width=.255\textwidth]{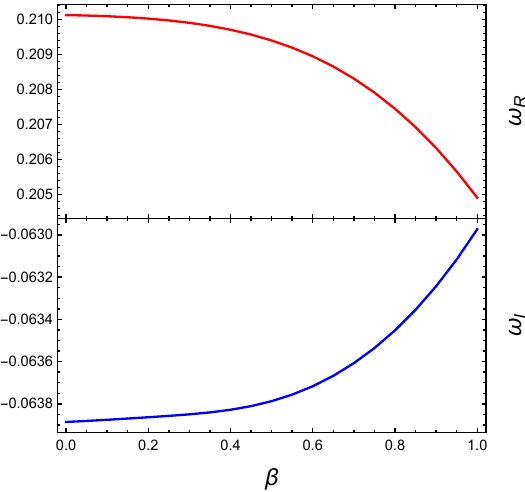}\\
\includegraphics[width=.375\textwidth]{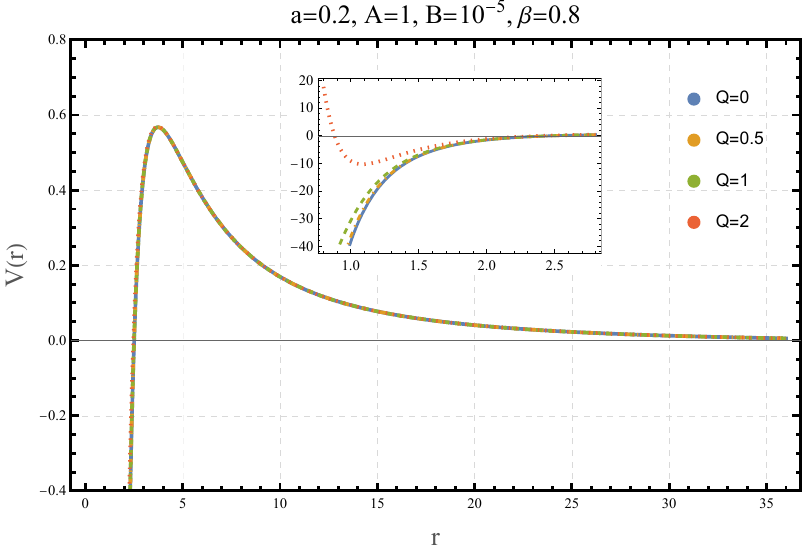}\hspace{0.03\textwidth}
\includegraphics[width=.23\textwidth]{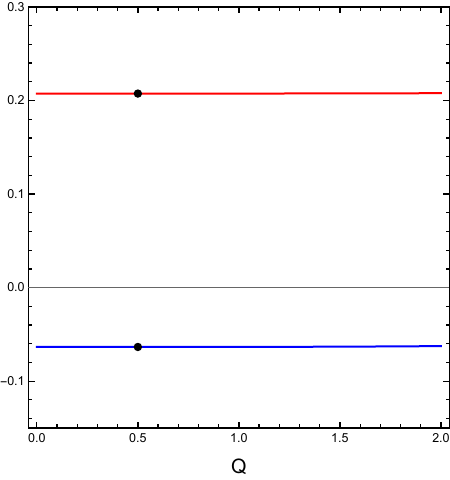}\hspace{0.03\textwidth}\includegraphics[width=.255\textwidth]{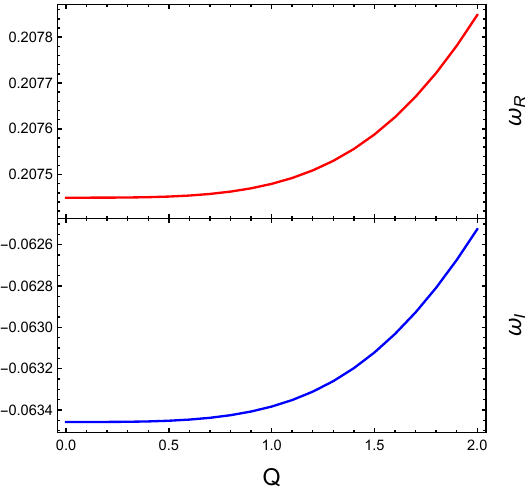}\\
\caption{\label{fig:qnm_metricdependence}
Evolution of the fundamental ($n=0$, $l=1$) QNMs under parameter variation. From left to right: modifications to the effective potential $V(r)$; the resulting frequency shifts presented on a fixed scale; and the same data on optimized scales to resolve detailed behaviors. The black dot in each panel indicates the reference MCDF-CoS configuration ($a=0.2$, $A=1$, $B=10^{-5}$, $\beta=0.8$, $Q=0.5$).}
\end{figure*}

\subsection{Computational methods}
\label{subsec:qnm_methods}

To compute the QNM frequencies, we employ two complementary semi-analytical techniques that are well-suited for effective potentials with a single maximum: the higher-order WKB approximation and the Mashhoon method.

The WKB method, originally adapted from quantum mechanics for black hole perturbation theory \cite{Schutz:1985km}, has been systematically developed to high orders \cite{Iyer:1986np, Konoplya:2003ii,Konoplya:2019hlu}. This technique matches asymptotic solutions across the potential's turning points via a Taylor expansion around its maximum $r_*^0$. The foundational third-order WKB formula is given by \cite{Iyer:1986np}:
\begin{equation}
\label{wkbexpension}
    \omega^2 = V_0 + \sqrt{-2 V_0^{(2)}} \, \Lambda_2(\mathcal{K}) - i \mathcal{K} \sqrt{-2 V_0^{(2)}} \, \left[ 1 + \Lambda_3(\mathcal{K}) \right],
\end{equation}
where $V_0$ is the maximum potential value, $V_0^{(2)}$ is its second derivative at $r_*^0$, $\mathcal{K} = n + 1/2$ is the overtone number, and $\Lambda_2$, $\Lambda_3$ are polynomials incorporating higher-order WKB corrections \cite{Iyer:1986np}. This method provides high accuracy for fundamental modes ($n=0$) with $l > n$, and its precision improves with increasing angular momentum $l$.

The Mashhoon method \cite{Ferrari:1984zz,Mashhoon:1982} offers an alternative, more intuitive approach by approximating the effective potential with the exactly solvable Pöschl-Teller potential:
\begin{equation}
    V_{\text{P-T}}(r_*) = \frac{V_0}{\cosh^2[\alpha (r_* - r_*^0)]},
\end{equation}
where the parameter $\alpha$ is determined by the curvature of the potential at its maximum: $\alpha^2 = -\frac{1}{2V_0} \frac{d^2V}{dr_*^2} \big|_{r_* = r_*^0}$. This approximation yields an analytic expression for the QNM frequencies:
\begin{equation}
    \omega_{\text{P-T}} = \sqrt{ V_0 - \frac{\alpha^2}{4} } - i \alpha \left( n + \frac{1}{2} \right).
\end{equation}
While generally less accurate than the higher-order WKB method for quantitative predictions, the Mashhoon method is highly effective for capturing qualitative trends and verifying parameter dependencies.

\subsection{Eikonal correspondence and shadow relationship}
\label{subsec:eikonal_correspondence}

In the eikonal limit ($l \gg 1$), a profound correspondence emerges between the wave dynamics of QNMs and the geometric properties of null geodesics \cite{Cardoso:2008bp}. In this regime, the effective potential \eqref{potential} simplifies to $V_{\text{eik}}(r) \approx f(r) l^2/r^2$ which relates directly to the null geodesic potential \eqref{geodesicpotential}, and its maximum coincides with the location of the unstable photon sphere. The QNM frequencies are then directly linked to the characteristics of null geodesics orbiting the photon sphere: $ \omega_{\text{eik}} \approx \omega_{\mathrm{PS}}$, where
\begin{equation}
\label{qnm_eik}
    \omega_{\mathrm{PS}} = \Omega_{\text{ps}} l - i \left( n + \frac{1}{2} \right) |\lambda|,
\end{equation}
with $\Omega_{\text{ps}}$ the angular frequency of photon orbits given by \eqref{angularvelocity}, and $\lambda$ the Lyapunov exponent, characterizing the instability timescale of these orbits, as defined in \eqref{Lyapunov}. This geometric correspondence extends further to the black hole shadow observed by a distant observer. The real part of the QNM frequency is inversely related to the shadow radius $R_s$: $\text{Re}(\omega_{\text{eik}}) \approx l \sqrt{f(r_O)} / R_s$, where $r_O$ is the observer's radial coordinate. This fundamental connection between the wave and geometric descriptions provides a robust cross-check for our computations and deepens the physical interpretation of both QNMs and the black hole shadow.

\subsection{QNM results and discussion}
\label{subsec:qnm_results}
 
\begin{figure*}[ht]
\centering % \begin{center}/\end{center} takes some additional vertical space
\includegraphics[width=.23\textwidth]{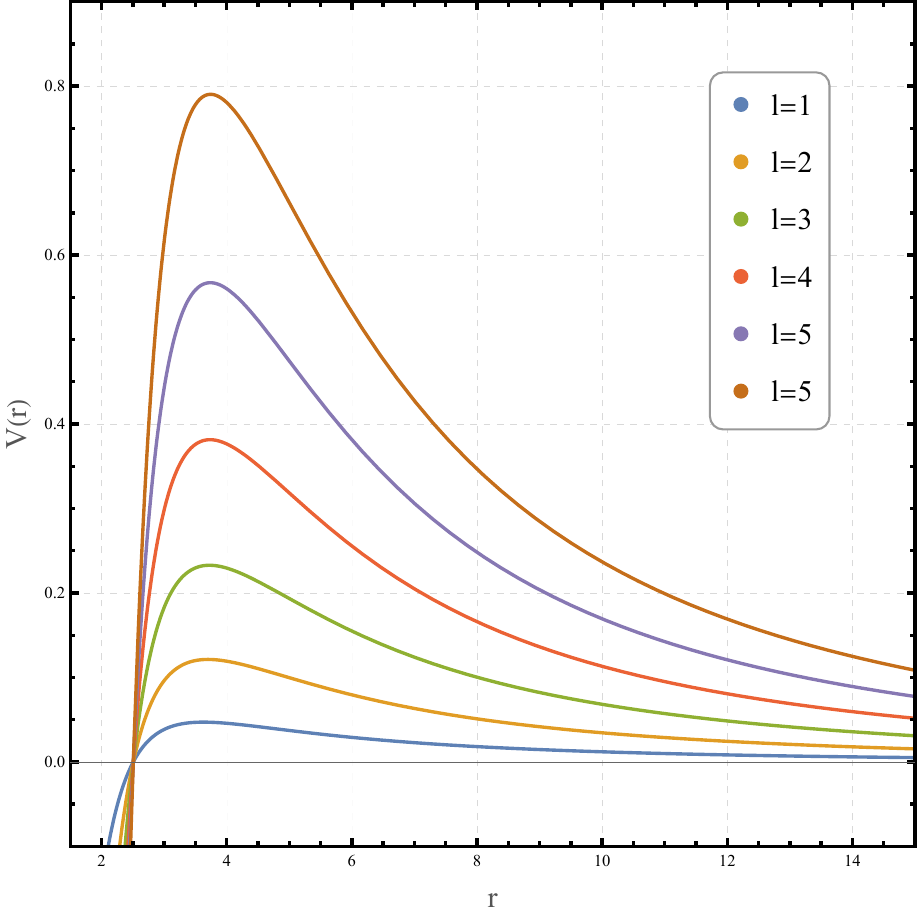}\hspace{0.02\textwidth}
\includegraphics[width=.33\textwidth]{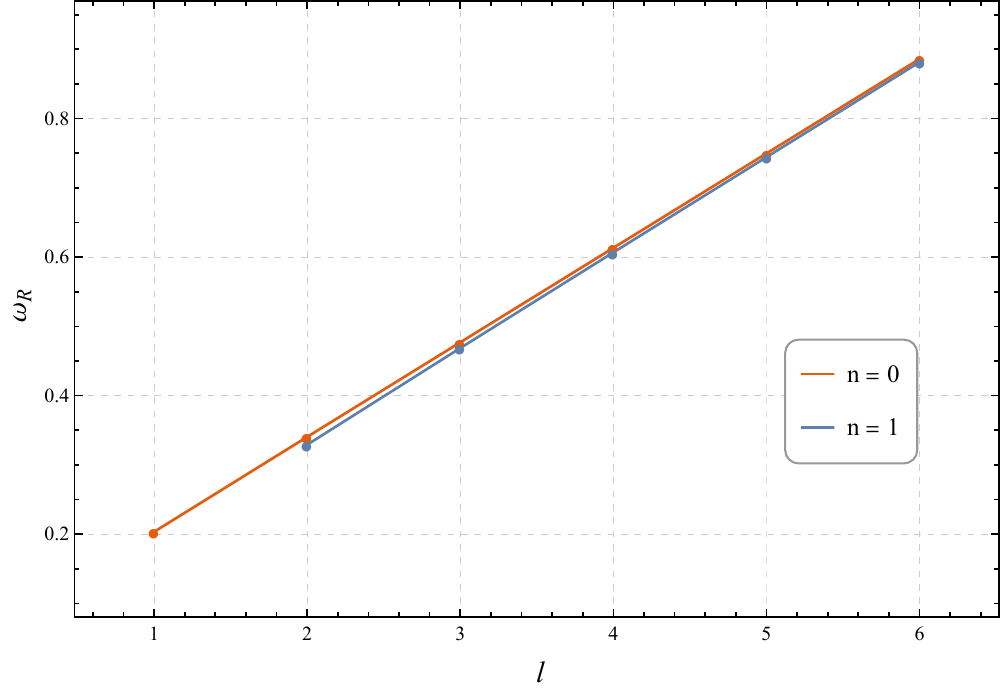}\hspace{0.02\textwidth}
\includegraphics[width=.33\textwidth]{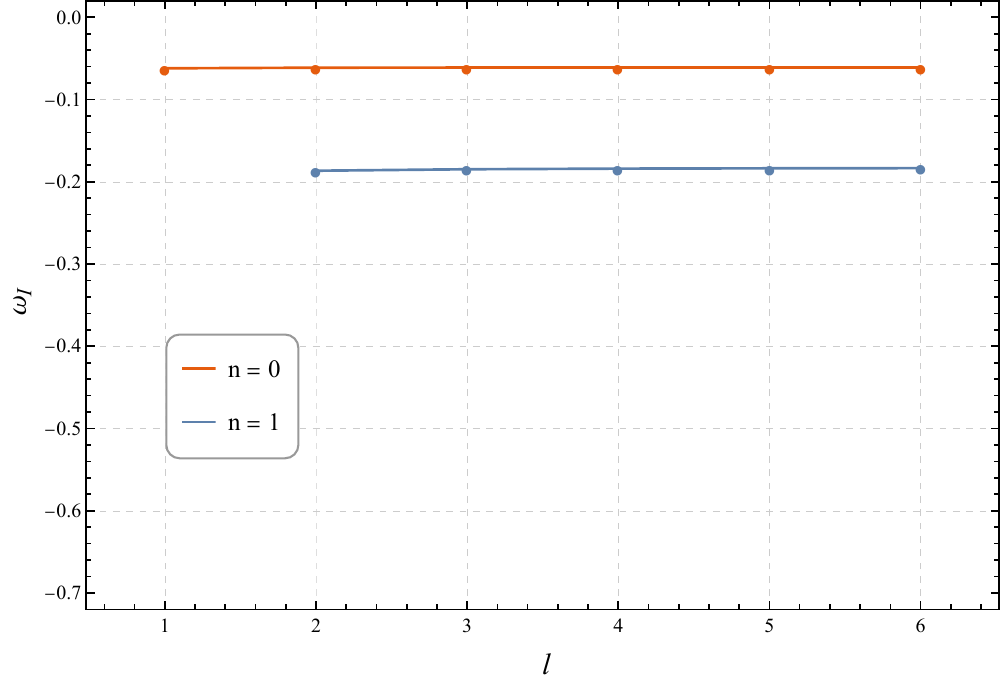}\\
\caption{\label{fig:qnm_ln} 
  Dependence of QNM frequencies on the angular quantum number $l$ and overtone number $n$ for the reference black hole configuration: $a=0.2$, $A=1$, $B=10^{-5}$, $\beta=0.8$, $Q=0.5$.}
\end{figure*}

Our systematic analysis of MCDF-CoS QNMs, computed via third-order WKB and Pöschl-Teller methods (Tables~\ref{tab:referencemodes}--\ref{tab:eikonal_limit}, Figs.~\ref{fig:referencemodes}--\ref{fig:qnm_ln}).
It is well established that the WKB approximation provides high accuracy for the fundamental modes ($n=0$), particularly in the eikonal regime ($l > n$), but its precision deteriorates rapidly for higher overtones ($n \ge 1$)~\cite{Iyer:1986np,Konoplya:2003ii}.
Consequently, we restrict our quantitative analysis to the fundamental mode ($n=0$) and the first overtone ($n=1$), omitting higher harmonics ($n \ge 2$) where the method becomes unreliable.
Generally, we observe strong correlations with the effective potential: higher barriers increase $\mathrm{Re}(\omega)$ (oscillation frequency), while broader barriers decrease $|\mathrm{Im}(\omega)|$ (damping rate).

The fundamental modes (\(n=0\), \(l=1\)) across different configurations (see Table~\ref{tab:referencemodes} and Fig.~\ref{fig:referencemodes}) exhibit high sensitivity to the matter content. The reference MCDF-CoS configuration shows substantially lower \(\mathrm{Re}(\omega)\) compared to the GCDF or CDF cases, along with notable variations in the damping rate. This indicates the predominant nontrivial effect of the MCDF parameter \(A\). 

Analysis of parameter dependence reveals that the CoS parameter \(a\) exerts the strongest influence (Table~\ref{tab:qnms_metricdependence}, Fig.~\ref{fig:qnm_metricdependence}): increasing \(a\) monotonically reduces both \(\mathrm{Re}(\omega)\) and \(|\mathrm{Im}(\omega)|\), leading to lower-frequency and longer-lived ringdown signals as a result of modifications in the spacetime geometry.

The MCDF parameters manifest more nuanced influences compared to the string cloud contribution (Table~\ref{tab:qnms_metricdependence}, Fig.~\ref{fig:qnm_metricdependence}). While the MCDF intensity parameter $Q$ and the polytropic index $\beta$ exhibit negligible impact across their physical ranges, the equation of state parameters $A$ and $B$ demonstrate significant, albeit distinct, modifications particularly at boundary values. As $A$ increases from its minimum allowed value, $\mathrm{Re}(\omega)$ undergoes an initial rapid decrease followed by convergence to a nearly constant value, with only marginal subsequent variation. Concurrently, $|\mathrm{Im}(\omega)|$ displays more constrained evolution within a narrow range, characterized by an initial decline succeeded by a slight enhancement before saturation. In contrast, increasing $B$ produces monotonic suppression of both oscillation frequency and damping rate—initially gradual, then accelerating, with a distinct upturn near the maximum allowed value of \(B\), ensuring the stability of the quasinormal modes by maintaining a positive oscillation frequency and a negative damping rate.

These spectral characteristics directly correspond to modifications of the effective potential barrier governing wave propagation. Parameters that enhance the potential barrier height (diminished $a$ or $B$) correspondingly increase $\mathrm{Re}(\omega)$ through stronger spatial confinement of wave modes. Conversely, parameters that broaden the potential barrier (enhanced $a$) reduce $|\mathrm{Im}(\omega)|$ by extending the interaction region and diminishing wave dissipation. The observed parameter-specific behavior near extremal values reveals a remarkable spatial decoupling: $A$ predominantly modulates physics in the near-horizon region, while $B$ governs modifications in the cosmological horizon domain. This spatial segregation underscores a fundamental correspondence between the MCDF equation of state parameters and the characteristic scales of the black hole spacetime, reflecting how the linear pressure component ($A$) and generalized Chaplygin term ($B$) dominate gravitational interactions in distinct regimes—the former influencing local strong-field dynamics, the latter controlling global cosmological evolution.

The overtone structure presented in Table~\ref{tab:qnm_LNdependence} and Fig.~\ref{fig:qnm_ln} compares the fundamental mode ($n=0$) with the first overtone ($n=1$).
Consistent with generic black hole behavior, the first overtone exhibits a larger $|\mathrm{Im}(\omega)|$ compared to the fundamental mode, indicating a faster decay rate.
For fundamental modes, $|\mathrm{Im}(\omega)|$ decreases and asymptotes with increasing $l$, while $\mathrm{Re}(\omega)$ grows approximately linearly.
%Comparison between the WKB and Pöschl-Teller results in Table~\ref{tab:qnm_LNdependence} shows good agreement for $n=0$, but the discrepancy increases for $n=1$, further justifying the exclusion of higher overtones from this semi-analytical treatment.

The eikonal limit verification (Table~\ref{tab:eikonal_limit}) demonstrates systematic convergence of WKB-computed frequencies $\omega_{\mathrm{WKB}}$ to the geometric prediction $\omega_{\mathrm{PS}}$, with relative errors $\Delta_R\%$, $\Delta_I\%$ decreasing with $l$. This confirms the photon sphere/QNM correspondence and underscores the fundamental wave-geometry relationship.

Throughout our analysis, the third-order WKB and Pöschl-Teller methods show excellent agreement for the fundamental modes ($n=0$), validating our computational approach in this regime.
However, the growing relative error observed for $n=1$ (as shown in Table~\ref{tab:qnm_LNdependence}) underscores the necessity of employing higher-precision numerical techniques, such as the continued-fraction method, should a detailed investigation of the higher-overtone spectrum be required in future studies.

\begin{table*}[]
	\centering
	\caption{Dependence of QNM frequencies on the angular quantum number $l$ and overtone number $n$ for the reference black hole configuration: $a=0.2$, $A=1$, $B=10^{-5}$, $\beta=0.8$, $Q=0.5$. The analysis is restricted to the fundamental mode ($n=0$) and the first overtone ($n=1$). $\Delta_R\%$ and $\Delta_I\%$ denote the relative errors for real and imaginary parts, calculated as $|\omega_{\mathrm{WKB}} - \omega_{\mathrm{P-T}}|/|\omega_{\mathrm{WKB}}| \times 100\%$.}
	\label{tab:qnm_LNdependence}
	\scriptsize
	\renewcommand{\arraystretch}{1.4}
	\setlength{\tabcolsep}{12pt}
	\begin{tabular}{c c c c c c}
		\toprule
		$l$ & $n$ & $\omega_{\mathrm{WKB}}$ & $\omega_{\mathrm{P-T}}$ & $\Delta_R\%$ & $\Delta_I\%$ \\
		\midrule
		1 & 0 & 0.203417 - 0.062029 $\mathrm{i}$ & 0.207452 - 0.063452 $\mathrm{i}$ & 1.96 & 2.29 \\
		\midrule
		2 & 0 & 0.340450 - 0.061319 $\mathrm{i}$ & 0.342732 - 0.061884 $\mathrm{i}$ & 0.67 & 0.92 \\
		2 & 1 & 0.329178 - 0.186588 $\mathrm{i}$ & 0.342732 - 0.185652 $\mathrm{i}$ & 4.11 & 0.50 \\
		\midrule
		3 & 0 & 0.476940 - 0.061138 $\mathrm{i}$ & 0.478537 - 0.061433 $\mathrm{i}$ & 0.33 & 0.48 \\
		3 & 1 & 0.468668 - 0.184780 $\mathrm{i}$ & 0.478537 - 0.184299 $\mathrm{i}$ & 2.10 & 0.26 \\
		\midrule
		4 & 0 & 0.613332 - 0.061066 $\mathrm{i}$ & 0.614562 - 0.061246 $\mathrm{i}$ & 0.20 & 0.29 \\
		4 & 1 & 0.606829 - 0.184029 $\mathrm{i}$ & 0.614562 - 0.183738 $\mathrm{i}$ & 1.27 & 0.16 \\
		\midrule
		5 & 0 & 0.749692 - 0.061030 $\mathrm{i}$ & 0.750694 - 0.061151 $\mathrm{i}$ & 0.13 & 0.20 \\
		5 & 1 & 0.744343 - 0.183647 $\mathrm{i}$ & 0.750694 - 0.183452 $\mathrm{i}$ & 0.85 & 0.11 \\
		\midrule
		6 & 0 & 0.886039 - 0.061009 $\mathrm{i}$ & 0.886885 - 0.061096 $\mathrm{i}$ & 0.10 & 0.14 \\
		6 & 1 & 0.881500 - 0.183428 $\mathrm{i}$ & 0.886885 - 0.183288 $\mathrm{i}$ & 0.61 & 0.08 \\
		\bottomrule
	\end{tabular}
\end{table*}

\begin{table*}[]
    \centering
    \caption{Verification of the eikonal limit correspondence: comparison between WKB-computed QNM frequencies ($\omega_{\mathrm{WKB}}$) and photon sphere-based predictions ($\omega_{\mathrm{PS}}$) for increasing angular quantum number $l$ (fundamental mode $n=0$). The relative percentage differences $\Delta_R\%$ and $\Delta_I\%$ demonstrate the convergence to the eikonal limit prediction as $l$ increases. Parameters: $a=0.2$, $A=1$, $B=10^{-5}$, $\beta=0.8$, $Q=0.5$.}
    \label{tab:eikonal_limit}
    \scriptsize
    \renewcommand{\arraystretch}{1.4}
    \setlength{\tabcolsep}{10pt}
    \begin{tabular}{c c c c c}
        \toprule
        $l$ & $\omega_{\mathrm{WKB}}$ & $\omega_{\mathrm{PS}}$ & $\Delta_R\%$ & $\Delta_I\%$ \\
        \midrule
        1    & 0.203417 - 0.062029 $\mi$   & 0.136327 - 0.060958 $\mi$   & 32.98        & 1.727        \\
        \midrule
        10   & 1.43138 - 0.060978 $\mi$    & 1.36327 - 0.060958 $\mi$    & 4.759        & 0.03223      \\
        \midrule
        100  & 13.7009 - 0.060958 $\mi$    & 13.6327 - 0.060958 $\mi$    & 0.4975       & $3.51\times10^{-4}$ \\
        \midrule
        300  & 40.9663 - 0.060958 $\mi$    & 40.8981 - 0.060958 $\mi$    & 0.1664       & $3.926\times10^{-5}$ \\
        \midrule
        500  & 68.2317 - 0.060958 $\mi$    & 68.1635 - 0.060958 $\mi$    & 0.0999       & $1.415\times10^{-5}$ \\
        \midrule
        $10^3$  & 136.3952 - 0.060958 $\mi$  & 136.327 - 0.060958 $\mi$    & $5.00\times10^{-2}$ & $3.542\times10^{-6}$ \\
        \midrule
        $10^4$  & 1363.339 - 0.060958 $\mi$  & 1363.27 - 0.060958 $\mi$    & $5.00\times10^{-3}$ & $3.545\times10^{-8}$ \\
        \bottomrule
    \end{tabular}
\end{table*}

\begin{figure*}[]
\centering % \begin{center}/\end{center} takes some additional vertical space
\includegraphics[width=.28\textwidth]{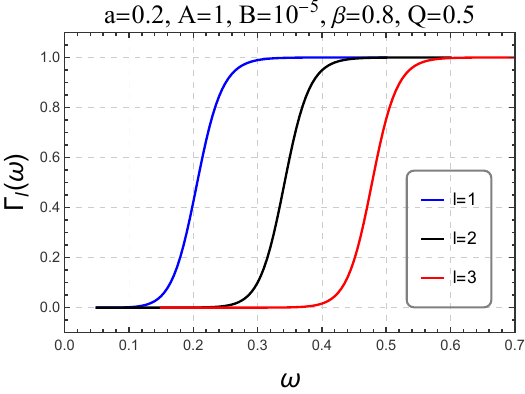}\hspace{0.03\textwidth}
\includegraphics[width=.28\textwidth]{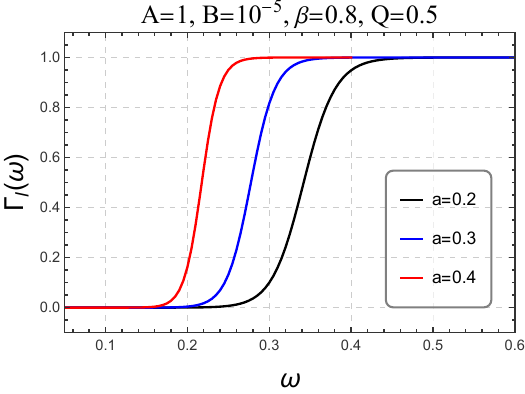}\hspace{0.03\textwidth}
\includegraphics[width=.28\textwidth]{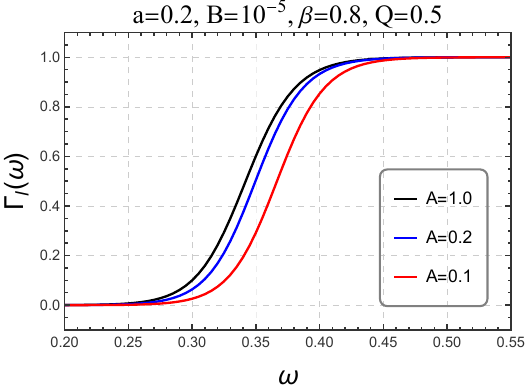}\\
\includegraphics[width=.28\textwidth]{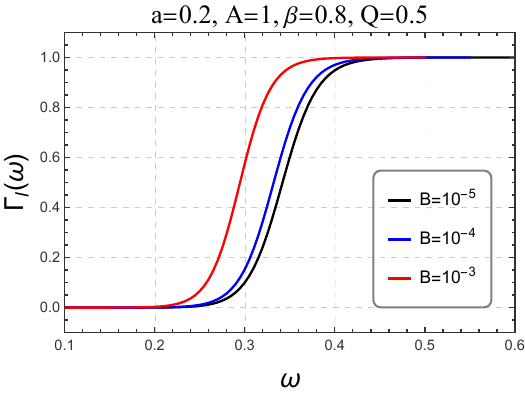}\hspace{0.03\textwidth}
\includegraphics[width=.28\textwidth]{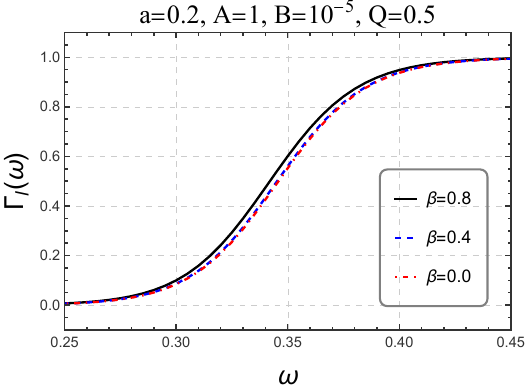}\hspace{0.03\textwidth}
\includegraphics[width=.28\textwidth]{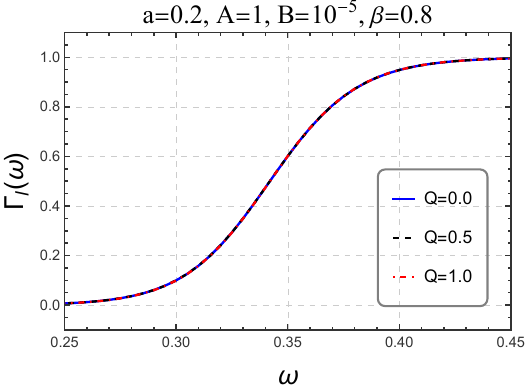}\\
\caption{\label{fig:gbf} 
Dependence of GBF spectrum $\Gamma_l(\omega)$ for MCDF-CoS black holes on the multipole number $l$ (the first panel) and on the spacetime parameters (subsequent panels where we have set $l=2$).}
\end{figure*}

\begin{figure*}[]
\centering % \begin{center}/\end{center} takes some additional vertical space
\includegraphics[width=.28\textwidth]{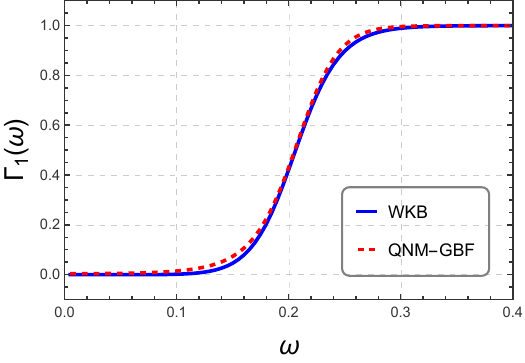}\hspace{0.03\textwidth}
\includegraphics[width=.28\textwidth]{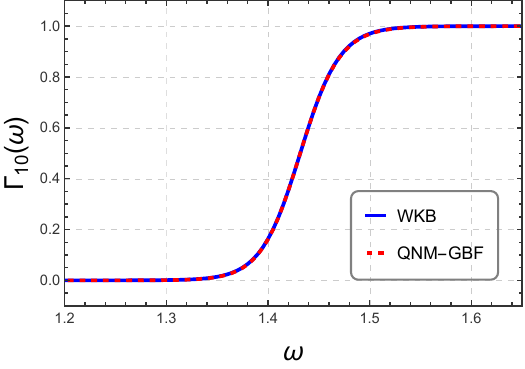}\hspace{0.03\textwidth}
\includegraphics[width=.28\textwidth]{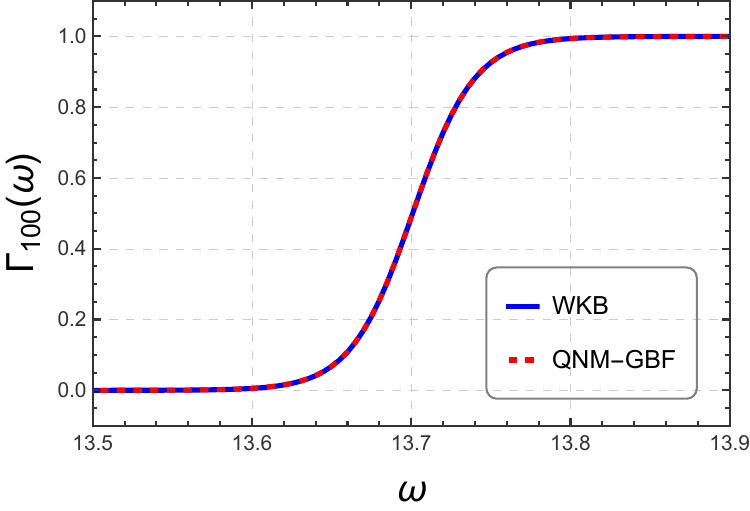}\\
\includegraphics[width=.28\textwidth]{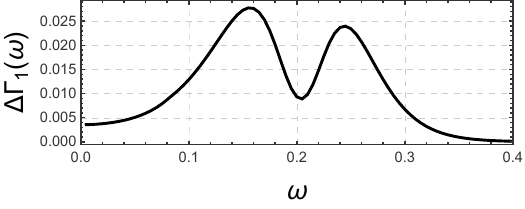}\hspace{0.03\textwidth}
\includegraphics[width=.28\textwidth]{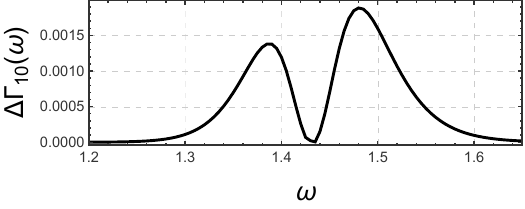}\hspace{0.03\textwidth}
\includegraphics[width=.28\textwidth]{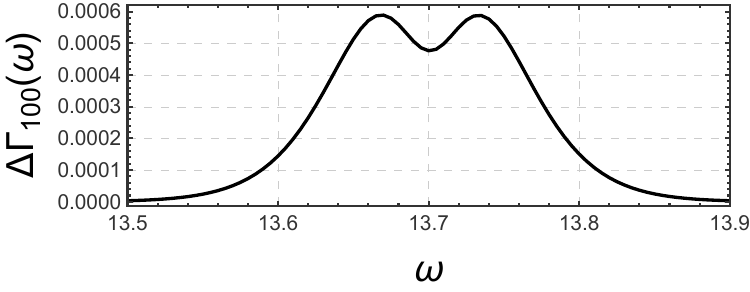}\\
\caption{\label{fig:gbf_correspondence} 
Comparison of greybody factors computed via the WKB method and the QNM correspondence. The left and center panels ($l=1, 10$) utilize the higher-order relation [Eq.~\eqref{K_function}] including overtone corrections, demonstrating significant improvement in accuracy. The right panel ($l=100$) retains the leading-order approximation, which remains asymptotically exact in the eikonal limit.}
\end{figure*}

\section{Greybody factors}
\label{sec:gbf}

Greybody factors (GBFs) quantify the transmission probability of Hawking radiation through the potential barrier to distant observers, crucially influencing the semi-classical evaporation process and observable radiation spectrum.

We consider massless scalar wave scattering governed by \eqref{waveeq}, but with scattering boundary conditions:
\begin{eqnarray}
   \Psi \sim T e^{-i\omega r_*} \quad (r_* \to -\infty), \\
   \Psi \sim e^{-i\omega r_*} + R e^{i\omega r_*} \quad (r_* \to +\infty),
\end{eqnarray}
where $T$ and $R$ are transmission and reflection coefficients for real frequency $\omega$. The GBF is defined as $\Gamma_l(\omega) = |T|^2 = 1 - |R|^2$.

We compute GBFs using the third-order WKB approximation \cite{Iyer:1986np, Konoplya:2019hlu}, which provides the semi-analytic formula:
\begin{equation}
	\label{gbf_expression}
    \Gamma_l(\omega) = \frac{1}{1 + \exp\left[ 2 \pi i \mathcal{K}(\omega) \right]},
\end{equation}
where $\mathcal{K}(\omega)$ is identical to the QNM function \eqref{wkbexpension} but evaluated for real $\omega$. This yields smooth transitions from total reflection ($\Gamma_l \approx 0$) to near-complete transmission ($\Gamma_l \approx 1$) as $\omega$ surpasses the potential barrier height.

\subsection{Eikonal limit connection}
\label{subsec:gbf_eikonal}

In the eikonal limit ($l \gg 1$), a profound correspondence emerges between GBFs, QNMs, and the geometric properties of the photon sphere \cite{Cardoso:2008bp,Konoplya:2024lir,Pedrotti:2025idg}. This relationship originates from the wave-optic equivalence principle in high-frequency regimes, where wave scattering is governed by the underlying null geodesic structure.

For massless scalar perturbations, the GBF can be expressed in terms of the quasinormal mode frequencies through the higher-order correspondence relation developed in Ref.~\cite{Konoplya:2024lir}. 
Utilizing the third-order WKB expansion, the exponent function $-i\mathcal{K}(\omega)$ in the GBF formula~\eqref{gbf_expression} is given by:
\begin{equation}
	\label{K_function}
	-i\mathcal{K}(\omega) = -\frac{\omega^2 - \mathrm{Re}(\omega_0)^2}{4\mathrm{Re}(\omega_0)\mathrm{Im}(\omega_0)} + \Delta_{\mathrm{HO}}(\omega, \omega_0, \omega_1).
\end{equation}
Here, $\omega_0$ and $\omega_1$ denote the frequencies of the fundamental mode and the first overtone, respectively. 
The higher-order correction term $\Delta_{\mathrm{HO}}$, which accounts for the anharmonicity of the effective potential, is explicitly defined in Ref.~\cite{Konoplya:2024lir} (see Eq.~(4.3) therein). 
This higher-order formulation significantly improves accuracy compared to the leading-order approximation, particularly in regimes where the potential deviates from a simple barrier shape.

Employing the eikonal QNM approximation \eqref{qnm_eik} for fundamental modes yields
\begin{equation}
\label{eikonal_qnm}
\omega_0 \approx \Omega_{\text{ps}} l - i\frac{|\lambda|}{2},
\end{equation}
with $\Omega_{\text{ps}}$ and $\lambda$ given by Eqs.~\eqref{angularvelocity} and \eqref{Lyapunov}, which leads to the geometric optics-based expression for the GBF \cite{Pedrotti:2025idg}:
\begin{equation}
\label{gbf_shadow_relation}
   \Gamma_l(\omega) \approx \frac{1}{1 + \exp\left[ -2\pi \left( \frac{\omega - \Omega_{\text{ps}} l}{|\lambda|} \right) \right]}.
\end{equation}

A notable observation is that while the higher-order QNM-GBF correspondence introduced in Eq.~\eqref{K_function} demonstrates excellent numerical agreement across the parameter space (Fig.~\ref{fig:gbf_correspondence}), the shadow-GBF relation in Eq.~\eqref{gbf_shadow_relation} exhibits more pronounced discrepancies. 
This enhanced sensitivity likely stems from the exponential amplification of small deviations between the actual QNM real part and its geometric optics approximation ($\mathrm{Re}(\omega_0) - \Omega_{\text{ps}} l$) within the GBF expression. 
Nevertheless, both formulations consistently underscore the fundamental role of photon sphere geometry in governing wave scattering dynamics.
Crucially, our analysis highlights a regime-dependent behavior in achieving this agreement: the inclusion of the higher-order correction term $\Delta_{\mathrm{HO}}$ is essential for low angular momentum modes (e.g., $l=1, 10$) where the effective potential is anharmonic, whereas the leading-order approximation naturally converges to the exact result in the eikonal limit (e.g., $l=100$) as the potential barrier becomes harmonic.

\subsection{GBF results and discussion}
\label{subsec:gbf_results}

The greybody factor spectrum $\Gamma_l(\omega)$ for MCDF-CoS black holes exhibits systematic dependencies on both angular momentum and spacetime parameters, with the transition frequency and profile sharpness serving as sensitive probes of the underlying geometry (Fig.~\ref{fig:gbf}).

Angular momentum dependence reveals that higher multipole numbers shift transmission thresholds to higher frequencies while simultaneously sharpening the transition profiles. For the reference configuration ($a=0.2$, $A=1$, $B=10^{-5}$, $\beta=0.8$, $Q=0.5$), the $l=1$ mode initiates significant transmission around $\omega \approx 0.1$, whereas the $l=3$ mode remains substantially suppressed until $\omega \approx 0.3$. This behavior is consistent with the increased potential barriers encountered by higher angular momentum states.

Among the spacetime parameters, the cloud of strings intensity $a$ exerts the most pronounced influence on the GBF spectrum. Increasing $a$ from 0.2 to 0.4 substantially lowers the transmission threshold—shifting the 50\% transmission point from $\omega \approx 0.25$ to $\omega \approx 0.15$ for $l=2$—while concurrently broadening the transition region. This modification reflects the softening of the effective potential barrier induced by the enhanced string cloud density.

The MCDF parameters $A$ and $B$ display distinctive modification patterns to the transmission spectrum, each governed by separate physical mechanisms. Parameter $B$ dominates near its maximum allowed value, where the intensified negative pressure component effectively lowers the potential barrier, substantially enhancing low-frequency transmission. Similarly, parameter $A$ produces its most significant effects at minimal values, where reduced fluid stiffness alters the potential profile, effectively enhancing the tunneling probability for low-frequency waves. This regime-dependent behavior underscores their different roles in the equation of state. In contrast, parameters $\beta$ and $Q$ exhibit negligible impact on the GBF spectrum across their physically allowed ranges, indicating their minimal influence on the scattering dynamics.

These systematic trends maintain consistent correlations with corresponding modifications in both the effective potential and QNM spectra discussed previously. The enhanced low-frequency transmission for larger $a$ values suggests potentially observable modifications to the Hawking radiation spectrum. The identified parameter sensitivities demonstrate that future high-precision measurements could constrain the string cloud intensity and key MCDF parameters ($A$ and $B$), while $\beta$ and $Q$ remain challenging to probe through GBF measurements alone.

\section{Summary and conclusions}
\label{conclusion}

In this comprehensive study, we have systematically investigated the gravitational properties of static, spherically symmetric black holes immersed in a modified Chaplygin-like dark fluid (MCDF) and a cloud of strings (CoS). By focusing on three key theoretical signatures---the black hole shadow, quasinormal modes (QNMs), and greybody factors (GBFs)---we have elucidated how these exotic environmental components collectively modify the fundamental characteristics and observable manifestations of black holes.

The spacetime geometry, characterized by the metric function ~\eqref{metric}, incorporates the MCDF through a hypergeometric function contribution \( G(r) \) and the CoS via the parameter \( a \). This framework encompasses several limiting cases and exhibits asymptotically de Sitter behavior, with the existence of black hole solutions imposing specific bounds on the parameter space.

Our analysis reveals a coherent and consistent picture across all three phenomena. The CoS parameter \( a \) emerges as the dominant factor, whose increase significantly enlarges the shadow radius, reduces both the oscillation frequency and damping rate of QNMs (leading to longer-lived ringdown signals), and substantially enhances low-frequency transmission in GBFs. This universal trend underscores the profound impact of the string cloud in globally softening the effective potential and altering the underlying spacetime geometry.

The influence of the MCDF parameters is more selective and reveals a nuanced hierarchy. The optical shadow is primarily affected at parameter extremes, where increasing \( A \) enlarges the shadow while increasing \( B \) reduces it. The QNM spectrum unveils a remarkable spatial decoupling: parameter \( A \), associated with the linear pressure component, predominantly modulates the near-horizon physics, whereas parameter \( B \), linked to the generalized Chaplygin term, governs the cosmological horizon domain. This spatial segregation is further reflected in the GBFs, where both parameters produce significant but regime-dependent effects: parameter \( B \) dominates near its maximum value by lowering the potential barrier, while parameter \( A \) exerts its strongest influence at minimal values by altering the potential profile to enhance low-frequency tunneling. Notably, the parameters \( Q \) and \( \beta \) exhibit negligible impact across all observational channels, indicating their subdominant role in the gravitational phenomenology studied here.

Beyond these parameter dependencies, our work highlights fundamental physical connections. The shadow appearance under different accretion flows demonstrates how accretion dynamics (e.g., Doppler beaming in the infalling model) shapes the morphology of the bright emission ring, independent of the shadow size itself---a purely geometric attribute. Furthermore, the verification of the eikonal limit firmly establishes the fundamental correspondence between the photon sphere (a geometric entity), the QNM spectrum (a wave entity), and the GBFs, thereby presenting a unified description of black hole dynamics that bridges wave and geometric optics.

Our findings suggest several promising avenues for future research. The identified parameter dependencies could inform efforts to constrain exotic matter components, particularly through shadow observations and QNM detections with current and next-generation instruments. The GBF results, while more theoretical at present, provide important insights into the scattering properties of black holes and would become directly relevant should evidence of Hawking radiation or other quantum emissions be discovered. Extending this analysis to rotating black holes would greatly enhance its astrophysical relevance.
Additionally, investigating electromagnetic perturbations and thermodynamic implications could provide a more complete characterization of these complex systems.

In conclusion, the MCDF-CoS black hole framework provides a rich theoretical laboratory for probing how exotic matter and extended structures modify fundamental black hole properties. The systematic dependencies and unified relationships we have identified not only deepen the theoretical understanding but also provide concrete, multi-faceted predictions for future astronomical observations aimed at deciphering the intricate environments surrounding these fascinating objects.

\section*{Acknowledgements}
This work was supported by the National Natural Science Foundation of China under Grant No. 12305070, and the Basic Research Program of Shanxi Province under Grant Nos. 202303021222018 and 202303021221033, and the China Scholarship Council under Grant No. 202308140133.

\bibliography{main}

\end{document}